\documentclass[conference,compsoc]{IEEEtran}
\IEEEoverridecommandlockouts
\usepackage{fancyhdr}
\fancypagestyle{preprint}{%
  \fancyhf{}
  \fancyhead[L]{\footnotesize\textsc{Preprint}}
  }

\usepackage{amsmath,amssymb}
\usepackage{graphicx}
\usepackage{booktabs}
\usepackage{multirow}
\usepackage{xspace}
\usepackage[table]{xcolor}
\usepackage[most]{tcolorbox}
\usepackage{cite}
\usepackage[colorlinks=true,
    linkcolor=blue,
    citecolor=blue,
    urlcolor=blue,
    filecolor=blue]{hyperref}
\usepackage{listings}
\usepackage{tabularx}
\usepackage{colortbl, rotating}
\usepackage{pdflscape}
\usepackage{fvextra}
\usepackage{tikz}
\usetikzlibrary{shadows.blur}
\usetikzlibrary{arrows.meta}
\usetikzlibrary{positioning}
\usepackage{adjustbox}
\usepackage{caption}

\definecolor{AttackBlueBG}{HTML}{DCE7F7}
\definecolor{OtherRedBG}{HTML}{FBE0DE}
\definecolor{genericquery}{RGB}{232,242,252}%
\definecolor{modelspecific}{RGB}{255,240,220}%

\newcommand{\eqnref}[1]{Eq.~\eqref{#1}}
\newcommand{\figref}[1]{Fig.~\ref{#1}}
\newcommand{\tabref}[1]{Tab.~\ref{#1}}
\newcommand{\secref}[1]{\S\ref{#1}}
\newcommand{\appref}[1]{App.~\ref{#1}}
\newcommand{\mainsection}[1]{\section{#1}}
\newcommand{\mainsubsection}[1]{\subsection{#1}}

\newcommand{\argsort}{\mathrm{argsort}}

\newcommand{\ours}{\texttt{TellTail}}
\newcommand{\oursrandom}{\texttt{TellTail-Random}}
\newcommand{\ourstopic}{\texttt{TellTail-Topic}}
\newcommand{\oursopt}{\texttt{TellTail-OPT}}

\newcommand{\unk}{\mathrm{UNK}}

\newcommand{\tmone}{\texttt{TM1}}
\newcommand{\tmtwo}{\texttt{TM2}}
\newcommand{\tmthree}{\texttt{TM3}}

\newcommand{\HP}{\emph{Harry Potter}}
\newcommand{\medicalqa}{MedicalQARetrieval}

\newcommand{\parhead}[1]{\vspace{0.5em}\noindent\textbf{#1.}\hspace{0.5em}}

\newtcolorbox{promptbox}[1][]{
  enhanced,
  breakable,
  colback=blue!3,
  colframe=blue!45!black,
  colbacktitle=blue!7,
  coltitle=blue!45!black,
  title=#1,
  fonttitle=\bfseries,
  boxrule=0.4pt,
  leftrule=2.2pt,
  rightrule=0.4pt,
  toprule=0.4pt,
  bottomrule=0.4pt,
  arc=1mm,
  left=2mm,
  right=2mm,
  top=1mm,
  bottom=1mm,
  before skip=6pt,
  after skip=6pt
}

\newtcolorbox{followupbox}[1][]{
  enhanced,
  breakable,
  colback=red!3,
  colframe=red!55!black,
  colbacktitle=red!6,
  coltitle=red!55!black,
  title=#1,
  fonttitle=\bfseries,
  boxrule=0.4pt,
  leftrule=2.2pt,
  rightrule=0.4pt,
  toprule=0.4pt,
  bottomrule=0.4pt,
  arc=1mm,
  left=2mm,
  right=2mm,
  top=1mm,
  bottom=1mm,
  before skip=6pt,
  after skip=6pt
}

\newtcolorbox{generationprompt}[1][]{
  enhanced,
  breakable,
  colback=blue!3,
  colframe=blue!45!black,
  colbacktitle=blue!7,
  coltitle=blue!45!black,
  title=#1,
  fonttitle=\bfseries,
  boxrule=0.4pt,
  leftrule=2.2pt,
  rightrule=0.4pt,
  toprule=0.4pt,
  bottomrule=0.4pt,
  arc=1mm,
  left=2mm,
  right=2mm,
  top=1mm,
  bottom=1mm,
  before skip=6pt,
  after skip=6pt
}

\newtcolorbox{ragprompt}[1][]{
  enhanced,
  breakable,
  colback=blue!3,
  colframe=blue!45!black,
  colbacktitle=blue!7,
  coltitle=blue!45!black,
  title=#1,
  fonttitle=\bfseries,
  boxrule=0.4pt,
  leftrule=2.2pt,
  rightrule=0.4pt,
  toprule=0.4pt,
  bottomrule=0.4pt,
  arc=1mm,
  left=2mm,
  right=2mm,
  top=1mm,
  bottom=1mm,
  before skip=6pt,
  after skip=6pt
}

\begin{document}

\title{The \ours{} of Embeddings: Fingerprinting Retrievers in Black-Box Systems}

\author{
\IEEEauthorblockN{Abdullah Garra$^{\dagger}$}
\IEEEauthorblockA{University of Massachusetts Amherst\\
agarrah@umass.edu}
\and
\IEEEauthorblockN{Matan Ben-Tov}
\IEEEauthorblockA{Tel Aviv University\\
matanbentov@mail.tau.ac.il}
\and
\IEEEauthorblockN{Mahmood Sharif}
\IEEEauthorblockA{Tel Aviv University\\
mahmoods@tauex.tau.ac.il}
}

\maketitle
\thispagestyle{preprint}
\begingroup
\renewcommand{\thefootnote}{$\dagger$}
\footnotetext{Work done while at Tel Aviv University.}
\endgroup

\begin{abstract}

Dense embedding models are core to modern text retrieval, enabling systems ranging from web search to retrieval-augmented generation (RAG).
Yet, retrievers are usually deployed within opaque systems, which expose only ranked results, cited sources, or generated answers.
This opacity prevents users from verifying which retrievers providers serve, and may create a false sense of robustness against retrieval attacks.
We show that attackers can infer retrievers' identity \textit{only} through queries.

We introduce \ours{}, a fingerprinting attack tailored for a spectrum of access levels to efficiently identify retrievers behind black-box systems.
Our insight is that, despite sharing many properties, retrievers can be steered to emit distinct results, demonstrated through two complementary strategies.
First, when the system exposes the retrieved passages, \ours{} compares retrieval overlap to deduce the underlying model.
Second,
when only the final generated response is exposed (e.g., in RAG),
\ours{} optimizes model-specific queries that induce a chosen retrieval behavior on the target retriever but transfer poorly to others.
Interestingly, the poor transferability that limits attacks elsewhere is exactly what makes them useful for fingerprinting.
We evaluate \ours{} using 53 retrievers under three increasingly restrictive settings, showing that retriever fingerprinting is feasible with a small query budget.
\ours{} perfectly identifies the deployed retriever from full retrieval rankings;
in 94.3\% of attempts when accessing unordered top-3 results;
and in 92.5\% of cases from language-model-generated answers alone.
Furthermore, we demonstrate that fingerprinting succeeds against a popular RAG system (OpenWebUI).
Overall, through exploiting model peculiarities,
we show that retrieval-based systems leak the identity of their embedding model---exposing intellectual property and enabling
reconnaissance ahead of model-specific attacks such as corpus poisoning.\footnote{We will make our code publicly available.}

\end{abstract}


\mainsection{Introduction}

Modern information-retrieval systems---web search, question answering, and retrieval-augmented generation (RAG)---that retrieve relevant passages from large text corpora
are commonly powered by \emph{dense retrieval}, in which a learned, neural embedding model maps both queries and passages into a shared vector space and ranks passages by their similarity to the query~\cite{karpukhin2020dpr,izacard2022contriever}.
The retrieved passages are then used differently across applications: returned directly as ranked results, shown as cited sources, or, in a RAG pipeline, used to enrich the context of a language model that generates the final answer~\cite{lewis2020rag}.
Yet these systems rarely reveal which embedding model underlies them~\cite{google-ai-overview,openai_projects,anthropic_projects}.

Identifying a hidden component of a system from its observable behavior is known as \emph{fingerprinting}, a useful primitive in security.
Fingerprinting schemes serve both benign and adversarial ends:
benign actors use it to verify and audit a system (e.g., confirming a provider serves a promised component),
while adversaries use it for reconnaissance ahead of component-specific attacks (e.g., fingerprinting a host's operating system to select an exploit; ~\cite{nmap_OS, anderson2017fingerprinting, greenwald2007toward}),
or to expose information a provider would rather hide, such as the intellectual property behind an application (e.g.,~\cite{azriel2017using}).
ML systems are no exception, with prior work showing that, e.g., carefully crafted queries can recover the exact version of an LLM behind an application~\cite{pasquini2025llmmap,zang2025errortrace,jiang2026ragid}.
Yet identifying the embedding model behind deployed retrieval systems remains unexplored.

This work shows that attackers can infer the deployed retriever \textit{only} by querying the system, whether it outputs retrieved results or merely generated answers.
At first glance, fingerprinting retrievers should be hard: they are trained on overlapping data, optimized with similar objectives, and their representations are argued to converge~\cite{huh2024platonic, jha2026harness}---suggesting retrievers should behave alike.
Yet we observe that they can be steered to emit distinct, model-specific%
results.
Building on this observation, we introduce \textbf{\ours{}}, a fingerprinting scheme that induces a desired retrieval pattern under one embedding model while transferring poorly to others.

At a high level, \ours{} attackers build a database of fingerprint queries for each
candidate retriever;
they then issue these queries to the target system
and match the observed behavior (e.g., the retrieved results, or the final response)
against the behavior expected under each candidate, predicting the most likely identity of the deployed retriever.

\begin{figure*}
    \centering
    \includegraphics[width=1\linewidth]{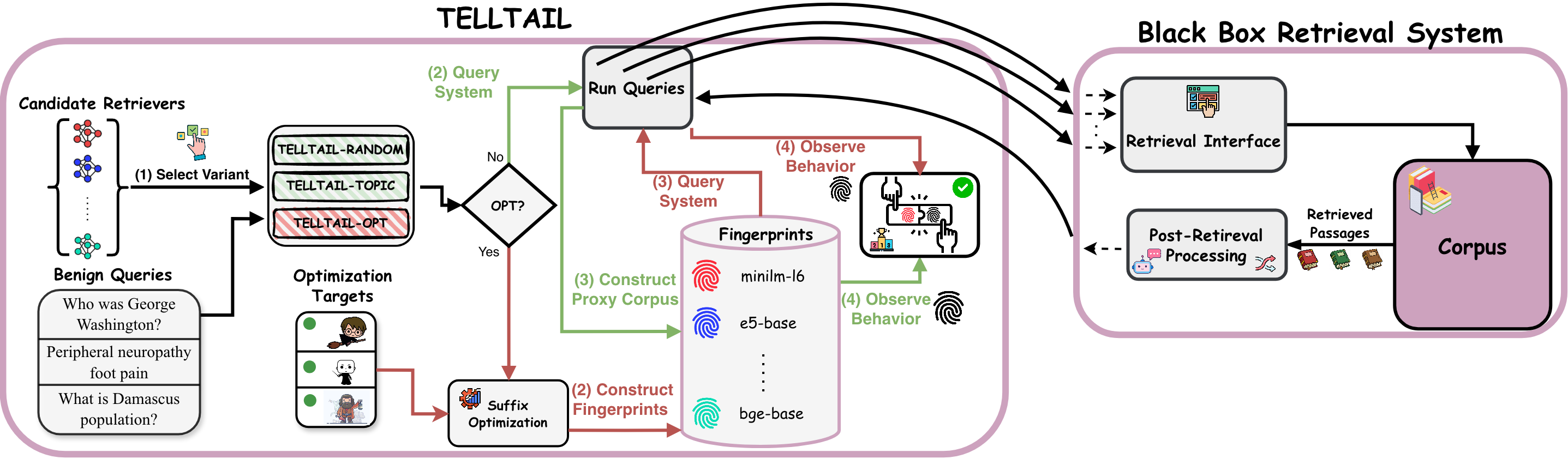}
    \caption{
    Overview of the fingerprinting procedure.
    Depending on the selected \ours{} variant, fingerprint construction follows one of two paths.
    For generic-query variants, \oursrandom{} and \ourstopic{} (green arrows), we first submit the benign queries
    to the target system, use the observed outputs to construct a proxy corpus for each candidate model,
    and then run the same queries against the proxy corpus
    to derive per-candidate sets of retrieved passages which act as fingerprints.
    The model-specific query variant, \oursopt{} (red arrows), optimizes model-specific query suffixes toward
    selected targets to construct these fingerprints.
    The resulting queries are then submitted to the black-box retrieval system,
    and the exposed behavior is
    compared with the candidate fingerprints to infer the deployed retriever.
    Depending on the available interface, the observed behavior may consist of
    retrieved passages or evidence from post-retrieval processing (e.g., LLM response).
    }
    \label{fig:telltail_overview}
\end{figure*}

The specific construction of the fingerprint queries and the expected
behavior depends on the fingerprinting strategy.
\ours{} is built on two efficient, complementary strategies spanning the access levels of retrieval systems, as illustrated in \figref{fig:telltail_overview}.
First, when retrieved passages are visible, \ours{} matches the target's retrieved passages of the fingerprint queries against those of each candidate model on a shared corpus (steps (2--4) of \oursrandom{} and \ourstopic{} in \figref{fig:telltail_overview}).
Second, when the passages are unavailable but inform the final generated response, \ours{} issues queries optimized to elicit behavior characteristic of one retriever while transferring poorly to others (steps (2--4) of \oursopt{} in \figref{fig:telltail_overview}).
Interestingly, such poor transferability---often a \textit{``bug''} limiting attack generalizability~\cite{liu2016delving,dong2018boosting,xie2019improving,wang2021enhancing}---here becomes a \textit{``feature''} serving the attacker: a query that succeeds only on the deployed model surfaces that model's particularities, revealing its identity.

We evaluate \ours{} against 53 retrievers under three increasingly restrictive retrieval interfaces, simulating an open-world setting in which the attacker has access to only 19 of the retrievers.
In the permissive setting, where full retrieval rankings are visible, \ours{} perfectly identifies the deployed retriever. When only unordered top-$3$ retrieved passages are visible, \ours{} identifies the correct retriever in 94.3\% of attempts. Even in the most restrictive setting, where retrieved passages are hidden and only generated answers are observed, \ours{} succeeds in 92.5\% of cases.
Finally, we demonstrate that \ours{} can fingerprint retrievers in a popular RAG system, OpenWebUI \cite{openwebui}, under realistic deployments.

Overall, our results show that retrieval-based systems leak the identity of their embedding model through observable retrieval behavior.
This leakage exposes intellectual property and supplies the reconnaissance that model-specific attacks tacitly assume: for instance, pernicious corpus-poisoning attacks that enable adversaries to increase the visibility of harmful content and manipulate RAG outputs rely on knowing the target retriever~\cite{zhong2023poisoning,zou2024poisonedrag,ben_tov2025gaslite}.
More broadly, \ours{} suggests that unexplained model idiosyncrasies are a practical, exploitable resource---fingerprinting being one of several ways an attacker might turn them to advantage.

In summary, we make the following contributions:
\begin{itemize}
\item We formulate the problem of fingerprinting retrievers behind black-box retrieval systems, across interfaces ranging from full rankings to answer-only responses (\secref{sec:threat_model}).
\item We introduce \ours{}, a scheme that exploits model-specific quirks to identify the retriever behind an opaque system (\secref{sec:method}).
\item Via extensive evaluation, we show retrievers can be reliably identified by querying the deployed system alone---even when it exposes only its generated answers (\secref{sec:results}).
\item We show that \ours{} succeeds on a popular RAG deployment, demonstrating its practicality (\secref{sec:case_study}).

\end{itemize}

\mainsection{Background and Related Work}
\label{sec:background}

\parhead{Dense retrieval}
Text retrieval ranks passages from a corpus by their relevance to a query.
Modern systems often implement this via \emph{dense retrieval}, where a neural \emph{embedding model} maps queries and passages into a shared vector space and ranks passages by vector similarity~\cite{karpukhin2020dpr,reimers2019sentencebert,izacard2022contriever}.
Concretely, in dense retrieval, a corpus of passages $\mathcal{P}=\{p_1,\dots,p_{|\mathcal{P}|}\}$ is encoded with a retriever $R$, whose embeddings are compared via a similarity measure $Sim(\cdot,\cdot)$ (e.g., cosine similarity).
Given a query $q$, the corpus is ranked by similarity to it:
\begin{equation}
\label{eq:ranking}
\pi_R(q) := \argsort\bigl(\{Sim(R(q),R(p)) \mid p\in \mathcal{P}\}\bigr),
\end{equation}
where $\pi_R(q)$ orders the passages by descending similarity to $q$ and $\pi_R(q)[:k]$ denotes the top-$k$ ranked passages.

\parhead{Retrieval-augmented systems}
Dense retrieval underlies a wide range of real-world applications, powering
semantic indexes in databases (e.g., Elasticsearch~\cite{elastic-vector-search}, Postgres \texttt{pgvector}~\cite{pgvector});
user-facing search (Meilisearch~\cite{meilisearch-github});
and, prominently, retrieval-augmented generation (RAG), where the retrieved passages enrich the context of a language model that produces the final answer~\cite{lewis2020rag} (e.g., Google's AI Overview~\cite{google-ai-overview}, NVIDIA ChatRTX~\cite{nvidia-chatrtx}, OpenWebUI~\cite{openwebui}).
We refer to the end-to-end pipeline---the retrieval module embedded within a larger application such as a search engine or a RAG assistant---as the \emph{retrieval-augmented system} (or, in short, \emph{retrieval systems}), distinguishing the deployed system from the retrieval method it uses internally.
These retrieval systems often do not expose which embedding model they use.
A user typically observes only artifacts of the interface's output---a ranked list of results, an unordered set of cited sources, or an LLM-generated answer---that obscures the underlying retriever's identity.

\parhead{Fingerprinting}
Inferring a hidden property of a system from its observable behavior is known as \emph{fingerprinting}, a side-channel attack and long-standing security primitive used, among others, to identify a host's operating system from network traffic~\cite{fyodor1998osdetect} or a user's browser from parameters it transmits and device characteristics~\cite{eckersley2010unique}.
Fingerprinting may serve both benign and malicious ends:
benign actors audit and verify systems (e.g., confirming a provider serves the component it advertises~\cite{wang2023publiccheck}),
while adversaries use it for reconnaissance ahead of component-specific attacks~\cite{nmap_OS, anderson2017fingerprinting, greenwald2007toward}, or to expose internals a provider would rather hide~\cite{azriel2017using}.

Prior work on fingerprinting also spans machine-learning models. These include identifying neural classifiers (e.g., by probing them%
with adversarial examples~\cite{cao2021ipguard}), but also generative models:
\textsc{LLMmap} recovers an LLM's exact version from%
crafted prompts~\cite{pasquini2025llmmap},
\textsc{ErrorTrace} attributes a model to its family via error patterns~\cite{zang2025errortrace}, and other work de-anonymizes text-to-image models behind voting-based leaderboards~\cite{naseh2026identifying}. Finally, RAG-ID detects whether a service is RAG-enhanced and enables identifying the LLM and knowledge base under black-box API access \cite{jiang2026ragid}.

Yet the ability to identify the \textit{embedding-based retriever model} behind a deployed retrieval system remains unexplored.
Identifying them may even seem unlikely, as prior work has argued that embedding models converge toward a shared
geometry~\cite{huh2024platonic,jha2026harness}, often retrieving overlapping passages for queries~\cite{caspari2024beyond}.
In this work, we study fingerprinting black-box retrieval systems, identifying their underlying embedding models merely by querying them.

\parhead{Attacks on retrievers}
A growing body of work attacks retrievers via \emph{corpus poisoning}, crafting adversarial passages that, once inserted into the corpus, are retrieved for a wide distribution of user queries~\cite{zhong2023poisoning,ben_tov2025gaslite}; such passages can subvert end-to-end RAG systems, steering the generated answer toward attacker-chosen content~\cite{zou2024poisonedrag,xue2024badrag}.
However, these attacks often rely on the attacker knowing the \textit{identity} of the deployed target model, tailoring the attack to it.
In this work, we show that an attacker can obtain this knowledge through an efficient fingerprinting scheme---\ours{}---acting as a reconnaissance stage and showing that a system gains little security from keeping its embedding model obscure.
Further, our work leverages these very attacks to optimize fingerprinting queries, as realized by the \oursopt{} variant (\secref{sec:fingerprinting_model_specific}).
\mainsection{Threat Model}
\label{sec:threat_model}

We consider an adversary aiming to identify, or \textit{fingerprint}, an underlying embedding-based retriever used within a black-box retrieval-augmented system.
As in classic systems-security fingerprinting (e.g., of an%
operating system~\cite{fyodor1998osdetect,nmap_OS}), the primitive has malicious implications,
spanning from intellectual property disclosure, to reconnaissance that informs subsequent corpus-poisoning attacks~\cite{zhong2023poisoning,ben_tov2025gaslite}.
However, fingerprinting may also have benign applications, such as verifying that a provider serves a promised model~\cite{wang2023publiccheck,ghodsi2017safetynets,kang2022scaling}.
Next, we describe the target retrieval system and the different interfaces it may pose, followed by the adversary goal and capabilities.

\mainsubsection{Target Retrieval System}
\label{sec:threat_model:target_sys}

We distinguish between the core \textit{retrieval module} and the broader \emph{retrieval-augmented system} (e.g., user-facing search or RAG) that wraps it.
The former is composed of a corpus $\mathcal{P}$ paired with a retrieval model $R$ that ranks passages by their similarity to a query, providing the top-$k$ passages (Eq.~\ref{eq:ranking}; $\pi_R(q)[:k]$).
The level of access an adversary has to retrieval results is an artifact of the retrieval system specific design.

\parhead{Retrieval Interfaces}
Users (or adversaries) interact with the retrieval module through a specific interface that governs the level of exposed retrieval information. We model this access via three distinct \textit{interfaces}:

\begin{itemize}
    \item \textbf{Full-Rank Interface (\tmone{}):} Given a query $q$, the system returns the \textit{full} ranking of passages $\pi_{R}(q)$ and the passages themselves. This reflects access to a semantic index in databases where the user can view both the passages and their ranking (e.g., search services \cite{meilisearch-github,elastic-vector-search}).

    \item \textbf{Unordered Top-$k$ Interface (\tmtwo{}):} Given a query $q$, the system reveals only the top-$k$ retrieved passages $\pi_{R}(q)[:k]$, \textit{not} necessarily ordered by similarity to the query. This is a standard setting for%
    retrieval APIs, and occasionally in RAG applications that list the retrieved results along with the answer, up to order changes~\cite{openwebui,nvidia-chatrtx,google-ai-overview}.

    \item \textbf{Response-only Interface (\tmthree{}):} Given a query $q$, the system returns a generated response (e.g., from an LLM) conditioned on the retrieved passages, without revealing the passages themselves. This is common in RAG deployments~\cite{lewis2020rag,openai_projects,anthropic_projects}.
    Here, the LLM output acts as a \emph{concept oracle}, implicitly leaking semantic properties of the retrieved content.
\end{itemize}

Additionally, modern retrieval systems may include preprocessing of the queries or passages (e.g., chunking or rephrasing \cite{retrieval-tricks--anthropic}), or supplement the%
retrieval stage with reranking \cite{retrieval-tricks--anthropic}.

\mainsubsection{The Adversary}
\label{sec:the_adversary}
The adversary's goal is to determine the retriever underlying the retrieval system.

\parhead{Objective}
We assume a realistic open-world setting where the adversary possesses a set of candidate retrievers $\mathcal{S}$ (e.g., popular open-source and proprietary models). The target retrieval-augmented system employs an unknown model $R^\star$ that may or may not be included in the adversary's candidate set. The adversary's goal is to determine if $R^\star \in \mathcal{S}$ and, if so, to identify precisely which model it is; otherwise, to state explicitly that the model is not in $\mathcal{S}$.

\parhead{Capabilities}
We assume the adversary has standard black-box query access to the target retrieval-augmented system, observing only the outputs permitted by the system's interface.
This interaction is constrained to a low query budget that is on order of  $|\mathcal{S}|$ (i.e., the number of candidate models).
The adversary has no knowledge of the system's internal specifications, weights, or full corpus;
however, the adversary often has some knowledge of the domain the system operates in (e.g., the specific aim or domain of the RAG application),
and adversaries can utilize visibility of top-ranked passages (as provided in \tmone{} and \tmtwo{}) to extract a small set of representative corpus passages or concepts (e.g., by issuing generic queries).
Finally, the adversary is assumed to have standard usage access to the candidate models $\in\mathcal{S}$, per each model's availability;
namely, when the model is accessible via public API (e.g., OpenAI's embedding models) attacker may compute embeddings, and when the model is open-source (e.g., available in HuggingFace) attacker may also have access to its weights, and may compute its gradients.
\mainsection{Methodology}
\label{sec:method}

\begin{table*}[t]
\centering
\caption{
Summary of the \ours{} strategies and variants. For each, the table presents how queries are constructed and the criteria used to identify the candidate retriever (ordered/unordered match on the top-$k$ responses from a proxy corpus, target passage appearing at top, or through an LLM response indicating the context is on an adversary-selected topic).
}
\label{tab:attack_configurations}

\resizebox{\textwidth}{!}{%
\begin{tabular}{
p{0.16\textwidth}
@{\hspace{8pt}}
p{0.13\textwidth}
@{\hspace{8pt}}
p{0.17\textwidth}
@{\hspace{8pt}}
p{0.22\textwidth}
@{\hspace{8pt}}
p{0.22\textwidth}
@{\hspace{8pt}}
p{0.26\textwidth}
}
\toprule

\textbf{Strategy}
& \textbf{Attack Variant}
& \textbf{Query Construction}
& \multicolumn{3}{c}{\textbf{Query-Match Criterion}} \\

\cmidrule(lr){4-6}

&
&
& \multicolumn{1}{c}{\textbf{\tmone{}}}
& \multicolumn{1}{c}{\textbf{\tmtwo{}}}
& \multicolumn{1}{c}{\textbf{\tmthree{}}} \\

\midrule

\rowcolor{genericquery}
& \textbf{\texttt{Random}}
& Sampled randomly
& \textit{Ordered} top-$k$ Match
& \textit{Unordered} top-$k$ Match
& N/A \\

\rowcolor{genericquery}
\multirow{-2}{*}{Generic queries}
& \textbf{\texttt{Topic}}
& Sampled from topic
& \textit{Ordered} top-$k$ Match
& \textit{Unordered} top-$k$ Match
& N/A \\

\cmidrule(lr){1-6}

\rowcolor{modelspecific}
Model-specific queries
& \textbf{\texttt{OPT}}
& Optimized per model
& Target appeared@$j\leq{}k$
& Target appeared@$k$
& LLM says all context is on target topic \\

\bottomrule
\end{tabular}%
}
\end{table*}

We now present \ours{}, a query-only fingerprinting attack for identifying retrievers in black-box retrieval systems, designed for a spectrum of interface access levels.%
\ours{} has two complementary strategies (summary in \tabref{tab:attack_configurations}).

The first strategy, \textit{generic-queries fingerprinting} (\secref{sec:generic_queries}), applies to the full-rank (\tmone{}) and unordered top-$k$ interfaces (\tmtwo{}), where retrieved passages are exposed to users.
This variant fingerprints the retriever by comparing the target system's retrieved passages to those of candidate retrievers on proxy corpus using \emph{generic} queries not specific to any model. We implement two variants of this strategy: \oursrandom{} and \ourstopic{}.

The second strategy uses \emph{per-model} optimized queries to induce model-specific retrieval behavior (\secref{sec:fingerprinting_model_specific}).
This strategy, implemented with \oursopt{}, is motivated by response-only interfaces (\tmthree{}), where retrieved passages are not directly observable and fingerprinting only
relies on the generated response.

Before detailing the \ours{} strategies and variants, we next formalize the fingerprinting mechanism (\secref{sec:adv_objective}).

\mainsubsection{Formalizing the Fingerprinting Mechanism}
\label{sec:adv_objective}
Let $S=\{R_1,\dots,R_m\}$ be a set of candidate%
retrievers, and let $R^\star$ be the unknown retriever deployed by the target system. The adversary issues $|Q|$ queries---reflecting their attack budget---denoted by:
$Q := \{q_1,\dots,q_{|Q|}\}$.

For each query $q_i$, the target retriever $R^\star$ induces a full ranking of the corpus, denoted by $\pi_{R^\star}(q_i)$. Under \tmone{} and \tmtwo{},for a query $q_i$, we consider the first $k$ retrieved passages $\pi_{R^\star}(q_i)[:k]$.
Under \tmone{}, each $\pi_{R^\star}(q_i)[:k]$ is observed as an \textit{ordered}
ranking, whereas under \tmtwo{} it is observed as an \textit{unordered} set.

Differently, under \tmthree{}, the attacker does not observe the retrieved passages. Instead, for a query $q_i$, they observe only the LLM-generated response conditioned on the top-$k$ retrieved passages,
$\mathrm{LLM}\!\left(\pi_{R^\star}(q_i)[:k]\right)$.
Across the query set $Q$, this yields $|Q|$ generated responses, one per query.

After receiving all $|Q|$ outputs, the adversary returns
$\hat{R} \in S \cup \{ \unk, \bot \}$
where $\unk$ indicates that the deployed retriever lies outside the candidate set, and $\bot$ denotes abstention due to insufficient confidence. The attacker, therefore, operates in an environment where they need to output $\unk$ when $R^\star \notin S$, and $\hat{R}=R^\star$ when $R^\star \in S$.

\mainsubsection{Fingerprinting With Generic Queries}
\label{sec:generic_queries}

We begin with fingerprinting using generic benign queries under \tmone{} and \tmtwo{}. These interfaces motivate a simple attack strategy: since the interface exposes retrieved passages, the attacker can first query the black-box system, which utilizes $R^\star$, with a set of queries $Q$ to construct a proxy corpus out of all retrieved passages,
$\bigcup_{i=1}^{|Q|} \pi_{R^\star}(q_i)[:k]$,
providing a common collection over which the retrieval behavior of the candidate retrievers can be compared to that observed from $R^\star$.
Once obtained, the attacker issues the same queries ($Q$) against this proxy corpus and performs retrieval each time using a different candidate retriever. The resulting retrieval results are compared to the passages exposed by the target ($R^\star$), and the candidate whose retrieved passages match exactly those of the $R^\star$ is identified as the target black-box system. The two concrete variants of fingerprinting with generic queries, \oursrandom{} and \ourstopic{}, differ only in the selection of queries ($Q$) they submit.

\parhead{\oursrandom{}} In the first variant,%
the attacker samples $|Q|$ queries uniformly from a broad distribution. This serves as a simple baseline: if candidate retrievers naturally disagree on ordinary queries, then their retrieval overlap with the target may already be sufficient for fingerprinting.

\parhead{\ourstopic{}} In the second variant, the attacker instead constructs $Q$ from a narrower semantic region, such as queries associated with the same topic, entity, or domain. For example, a topic-focused query set might contain \emph{``George Washington''} and \emph{``The first president of the US''} whereas a random set might contain unrelated queries such as \emph{``What is the capital of France?''} and \emph{``What causes a thunderstorm?''}. The former queries are likely to retrieve semantically related passages, producing a less diverse proxy corpus with multiple plausible passages relevant to both queries. In contrast, random queries may retrieve distinct and relatively unambiguous passages on which many embedding models agree. By concentrating queries in a narrow semantic region, candidate retrievers are therefore more likely to differ in how they rank and select among similar relevant passages, amplifying retrieval differences and making the observed output more informative about the deployed retriever's identity.

\parhead{Scoring}
Given both the observed retrieved passages w.r.t. the target retrieval system ($\pi_{R^\star}(q_i)[:k]$), and the retrieved passages w.r.t. the proxy corpus ($\hat{\pi}_R(q_i)[:k]$), we say a candidate retriever \textit{matches} the system's retriever identity on fingerprint query $q$, if both of the query's retrieved passages agree on their top-k results.
For \tmone{} this \textit{match} requires the ordered top retrieved passages to be identical, while for TM2 compares the unordered set of the top retrieved passages.
For a fingerprint scheme using a set queries $Q$, we define the \textit{acceptance score} of the candidate retriever as the target retriever, as the fraction of queries in $Q$ yielding a match.

\parhead{Attack Budget} Because the same query set $Q$ is shared across all candidate
retrievers, the generic-query variants require $|Q|$ queries to the
target system, independently of the candidate-set size $|S|$.

\mainsubsection{Fingerprinting With Model-specific Queries}
\label{sec:fingerprinting_model_specific}

\begin{figure}[t]
    \centering
    \includegraphics[width=1\columnwidth]{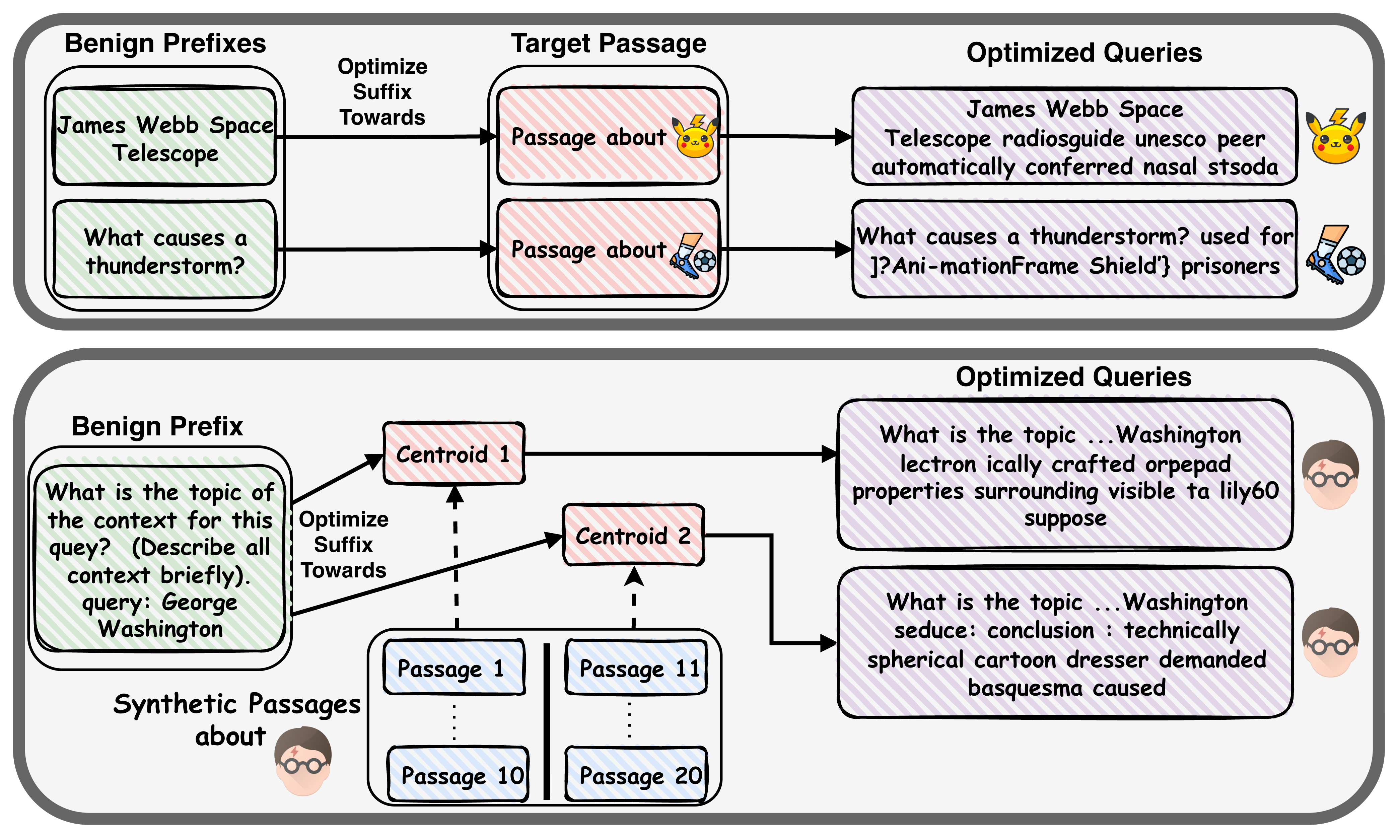}
    \caption{Overview of the \oursopt{} method. When passages are available, \oursopt{} optimizes benign queries towards specific target passages create the model-specific queries. If no passages are exposed by the system, \oursopt{} creates multiple queries, each is optimized towards a centroid of passages representing the topic.}
    \label{fig:telltail_opt_overview}
\end{figure}

\parhead{\oursopt{}} Unlike \oursrandom{} and \ourstopic{}, which rely on naturally occurring differences between models, \oursopt{} deliberately constructs model-specific fingerprints. In this variant, for each candidate retriever, we optimize a set of queries designed to induce a predetermined retrieval behavior (e.g., promote a specific passage) when that retriever is deployed by the target system. We focus on suffix-based optimization and adapt techniques from corpus-poisoning attacks, GASLITE, and its closed-weights extension RASLITE~\cite{ben_tov2025gaslite,ben2026tropt}.
These methods optimize adversarial suffixes appended to objects in order to shift their embeddings toward a desired objective in representation space. In our case, given some benign query prefix, $q_{\mathrm{pre}}$, this optimization attempts to find a suffix $s$ that maximizes the similarity between the embedding of the concatenated query $q_{\mathrm{pre}} \Vert s$ and a target vector $z_R$ under a candidate retriever $R$. The optimization target $z_R$ can represent the embedding of a specific passage or more generally a semantic direction corresponding to a target topic. In addition, $z_R$ is chosen to be semantically unrelated to the benign query prefix $q_{\mathrm{pre}}$, such that the corresponding passage or topic would not ordinarily be retrieved for $q_{\mathrm{pre}}$ alone (e.g., a prefix concerning \emph{George Washington} may be optimized toward a passage about \HP{}). Hence, the optimization objective is:

\begin{equation}
\label{eq:optimized_suffix_objective}
s_R^\star(q_{\mathrm{pre}})
=
\arg\max_{s \in \mathcal{T}_R^\ell}
Sim\left(
R(q_{\mathrm{pre}} \Vert s), z_R
\right).
\end{equation}

where $\mathcal{T}_R^\ell$ denotes the set of valid length-$\ell$ textual suffixes considered by the optimizer, subject to further constraints we introduce below. \figref{fig:telltail_opt_overview} illustrates \oursopt{}.

\parhead{Passage-level Optimization}
Under \tmone{} and \tmtwo{}, \oursopt{} observes passage-level retrieval outputs and can%
optimize toward a specific retrieved passage. Concretely, for a target passage $p \in \mathcal{P}$, we set the optimization target in \eqnref{eq:optimized_suffix_objective} to its embedding%
$z = R(p)$ such that the optimized query promotes $p$ in candidate retriever $R$'s ranking.

\parhead{Topic-level Optimization}
Under \tmthree{}, the attacker observes only the LLM-generated answer rather than the retrieved passages themselves. As a result, promoting a single passage to the LLM's relevant context may have little visible effect on the final response. Instead, we optimize toward a broader target topic with the goal of making topic-related passages sufficiently prevalent in the retrieved context to influence the LLM’s answer. To this end, we instantiate the query prefix \(q_{\mathrm{pre}}\) in Eq.~\ref{eq:optimized_suffix_objective} as a concatenation of a generic request to describe the retrieved context and an otherwise benign query semantically unrelated to the topic the suffix is optimized toward
(i.e., $q_{pre}=describe\_context || misc\_benign\_query$).%

Following prior work that represents a semantic topic by averaging
embeddings of passages associated with that topic
\cite{ben_tov2025gaslite}, we define the target vector as the centroid of
a set of target-topic passages. Given such a passage set
$\hat{\mathcal{P}}$, which can be collected in several ways (e.g., LLM-generated passages or book excerpts), we define $z_R$ as:
\[
z_R = \frac{1}{|\hat{\mathcal{P}}|}\sum_{p \in \hat{\mathcal{P}}} R(p).
\]

\parhead{Limiting Transferability With Token Blocking}
In many adversarial settings, transferability is desirable~\cite{liu2016delving,dong2018boosting,xie2019improving,wang2021enhancing}: an attack is
stronger if it succeeds across many models. In fingerprinting, however,
transferability is counterproductive. A query optimized for one retriever should
induce the intended behavior under that retriever, but not under other
candidate retrievers.
Empirically, we find that unconstrained suffix optimization of \eqnref{eq:optimized_suffix_objective}
often violates this requirement and yields optimized queries that
transfer broadly across retrievers (e.g., promote the target passage for multiple retrievers).
Concretely, we observe that optimized suffixes tend to use terms that either appear directly in the content being promoted or are semantically related to it. Moreover, blocking only exact tokens is insufficient, as semantically related terms (e.g., synonyms) can still induce transferability across retrievers.

We therefore use a token-blocking mechanism that excludes both exact and semantically related tokens from the optimizer's search space. We first block tokens that appear directly in the content defining the optimization target. For passage-level optimization, this consists of all token IDs appearing in the target passage $p$. For topic-level optimization, we instead block tokens appearing in the passages used to construct the topic centroid $z_R$.

We then semantically expand this initial blocked set up to a fixed blocking budget. Consider first passage-level optimization. For open-weight retrievers, each vocabulary token is scored by its maximum cosine similarity to the contextualized token representations of the target passage, and the highest-scoring vocabulary tokens are added to the blocked set. For closed-weight retrievers, where contextualized token representations are unavailable, we use the embedding API as a surrogate: we independently embed vocabulary tokens and individual words from the target passage, and score each vocabulary token by its maximum similarity to these word embeddings. For topic-level optimization, we apply the same procedure by aggregating the token-level similarity scores across all passages used to construct the topic centroid. Thus, in both cases, the blocked set contains tokens appearing directly in the relevant passages and also vocabulary that is semantically close to their content. We further quantify the effect of token blocking in \appref{app:blocking_effects}.

\parhead{Scoring}
The \textit{match} criterion for \oursopt{} depends on the target interface.
In
\tmone{} and \tmtwo{}, each optimized query is constructed to promote a
specific target passage. After issuing the query to the target system,
we check whether that passage appears in the observed retrieval output (e.g., top-$k$).

Under \tmone{}, the attacker observes the ranking and can therefore
restrict scoring to the highest-ranked retrieved passages.
We use appeared@$j$, $j \leq k$: a query is counted as a match only
when the target passage appears among the top-$j$ retrieved passages.
Because rank information is available in this setting, the attacker can focus
on a stronger signal provided by higher-ranked results rather than using the entire exposed top-$k$ list. In our experiments, we set $j=\min(k,3)$, i.e., we use appeared@$3$ unless fewer ranks are available.
We found limiting to comparison to the higher-ranked passages to yield better fingerprinting (\secref{sec:eval_tm2}).

Under \tmtwo{}, the attacker observes the top-$k$ passages without order. Here, we score using appeared@$k$: a
query is counted as a match when the target passage appears anywhere in
the observed top-$k$ set.

In \tmthree{} on the other hand, the retrieved passages are unavailable, and the per-query match criterion is defined over the generated answer. Concretely, the attacker inspects whether the answer to the optimized fingerprint prompt completely relates to the target topic, and match the retrievers accordingly.

As with the generic-query variants (\secref{sec:generic_queries}), we define the \textit{acceptance score} of a candidate retriever as the fraction of its optimized queries yielding a match.

\parhead{Attack Budget} As a separate query set is optimized per
candidate, the number of model-specific queries is
\[
\sum_{R\in S}|Q_R|.
\]
When all candidate-specific query sets have the same size, this becomes
$|Q_R|\cdot|S|$ and therefore scales linearly with the candidate-set
size. The complete online query budget $Q$ may additionally include
interface-specific exploratory queries, as detailed in
\secref{sec:setup_attack_instantiation}.

\mainsection{Experimental Setup}
\label{sec:experimental_setup}

We evaluate our attack variants across three different threat models and various retrievers in a trial-based setup.
In each trial, one retriever $R^\star$ from our evaluated model set is used as the underlying retrieval model of the target retrieval-augmented system.
The adversary has access to a candidate set of retrievers $S$, issues a number of queries to the target system,  observes only the information exposed by the corresponding interface, and predicts the underlying retriever.

Our systematic evaluation considers three retrieval interfaces
(\secref{sec:threat_model}). Under \tmone{} and \tmtwo{} the system returns passages. For \tmthree{}, we evaluate a
controlled vanilla RAG pipeline: the target retriever returns the most relevant
passages from our corpus,%
which are then inserted into a fixed
RAG prompt to query an LLM.%
(\appref{app:RAG_prompt} presents the RAG prompt.)
Since this setup isolates
the retriever's effect on the generated response by excluding additional
deployment components (e.g., query rephrasing and reranking), we extend our evaluation under \tmthree{} with a separate
end-to-end case study in OpenWebUI \cite{openwebui} in \secref{sec:case_study}.

\mainsubsection{Models and Corpus}
\label{sec:setup_models_data}

\parhead{Embedding Models}
We evaluate \ours{} on a pool of $53$ embedding models, each form a single trial.
The pool covers
a diverse set of retriever families, model sizes,
training objectives, and deployment regimes.
It includes popular and high-performing
retrievers from standard embedding benchmarks such as MTEB
\cite{muennighoff2023mteb}, both open-source and proprietary API-only models,
along with multiple models of the \textit{same} family---to test the sensitivity of fingerprinting methods.
Out of the $53$ models, we select a diverse candidate set containing 19 models, spanning
multiple families, sizes, and backbone designs, including closely related
variants that make fine-grained fingerprinting more challenging.
The remaining $34$ are retrievers that are not accessed by the attacker in this evaluation
and are therefore expected to be classified as \textit{unknown} models.
The full list of $53$ models is provided in \appref{app:models};
the candidate set $S$ of 19 retrievers consists of:
\emph{qwen-0.6b} and \emph{qwen-4b} \cite{zhang2025qwen3};
\emph{e5-small}, \emph{e5-base}, and \emph{e5-large} \cite{wang2024E5};
\emph{gte-modernbert}, \emph{gte-large-v1.5}, \emph{gte-base-v1.5}, and \emph{gte-multi} \cite{zhang2024mgte};
\emph{nomic-modernbert} and \emph{nomic-v1.5} \cite{nussbaum2024nomic};
\emph{gtr-t5-base} \cite{ni2022gtr};
\emph{minilm-l6} and \emph{minilm-l12} \cite{wang2020minilm};
\emph{mpnet} \cite{song2020mpnet};
\emph{snowflake-v2} \cite{merrick2405arctic};
\emph{gemma-300m} \cite{vera2025embeddinggemma};
\emph{bge-base} \cite{bge_embedding}; and the closed-weights model \emph{text-embedding-3-small} (OpenAI) \cite{openai2024textembedding3small}.

\parhead{Corpus}
As our primary dataset, we use the MS MARCO passage corpus as the
retrieval corpus~\cite{bajaj2016ms}. MS MARCO contains roughly
$8.8$ million passages collected from web documents associated with
real search queries, making it a large and semantically diverse retrieval
benchmark.
We further validate our findings on \medicalqa{} \cite{muennighoff2023mteb,mmteb,BenAbacha-BMC-2019}, a small, semantically dense dataset comprising approximately 2.4K passages focused on medical topics (the results, consistent with MS MARCO's, are in \appref{app:medicalqa_evaluation}).%
We defer additional implementation details, including corpus indexing, to \appref{corpus_indexing}.

\mainsubsection{Attack Instantiation}
\label{sec:setup_attack_instantiation}

\parhead{Instantiations in \tmone{} and \tmtwo{}}
For \tmone{} and \tmtwo{}, we instantiate the \textit{generic-query}
variants with a shared fingerprint query set $Q$ of size $|Q|=20$.
\oursrandom{} uses $20$ random queries spanning a broad range of
unrelated topics. \ourstopic{} uses the same number of queries, but all
queries concern a single semantic topic, \HP{}, yielding a
more localized query set. The complete query sets used by both variants
are provided in \appref{app:queries_prefixes} (\tabref{tab:telltail_generic_queries}). We additionally evaluate \ourstopic{} on a different topic (\emph{neurological disorders}) within \medicalqa{} in \appref{app:medicalqa_evaluation}, showing that its effectiveness is not specific to a particular choice of topic.

For \oursopt{}, as illustrated in \figref{fig:telltail_opt_overview}, since passages are visible under \tmone{} and \tmtwo{}, we first use an LLM to generate $|Q_R|=20$ random exploratory queries as benign prefixes ($q_{pre}$) from a broad range of topics.
The queries are used to collect%
passages from the system's knowledge base.
Then, we pair each $q_{pre}$ to an unrelated retrieved passage, and use it for the fingerprint-query optimization.
The concrete query prefixes used $Q_R$ can be found in~\appref{app:queries_prefixes} (\tabref{tab:gaslite_prefix_queries}).
Next, for each candidate retriever $R\in S$ and query prefix, we optimize a model-specific
suffix toward the paired target passage using the GASLITE \cite{ben_tov2025gaslite} (and, for black-box retrievers, RASLITE \cite{ben2026tropt})
optimization procedure. The token-blocking
mechanism follows
\secref{sec:fingerprinting_model_specific}.
The attack therefore first submits to the system $|Q_R|$ benign queries to
obtain the target passages, followed by $|Q_R|$ optimized queries per
candidate retriever, yielding a total attack budget ($|Q|$) of $|Q_R| + |Q_R|\cdot|S|$.
In our experiments, $|Q_R|=20$ and $|S|=19$, yielding a total of
$400$ queries to the target system.

The same queries, optimized suffixes, and target passages are used under
\tmone{} and \tmtwo{}.
We also evaluate different observation cutoffs \(k\), corresponding to the number of passages exposed under \tmone{} and \tmtwo{} (top-$k$), ranging from \(1\) to \(50\).

\parhead{\tmthree{} Instantiation}
To simulate the target system we use OpenWebUI, a common RAG system, which, given a query, retrieves the top-$3$ passages, and prompts the \texttt{gpt-4o-mini} LLM \cite{openai2024gpt4omini} to generate a response.

Under this%
response-only interface, we perform the topic-level optimization procedure described in~\secref{sec:fingerprinting_model_specific}, designed to steer the retrieved passages' topic and the LLM response accordingly.
Per the attacker familiarity with the target system (\secref{sec:threat_model}), we assume it holds knowledge of a particular topic present in the corpus. In our case this chosen topic is \HP{}.
The attack first defines a shared benign query prefix, which instructs the LLM to describe
the retrieved contextand includes an unrelated query topic (e.g., \emph{George Washington}).
Then, using GASLITE (and RASLITE) we optimize 10 suffixes appended to this query: each suffix is optimized towards the center of $10$ synthetic, LLM-generated passages containing diverse facts about or scene-descriptions from \HP{}.
The attacker then inspects the generated responses using an LLM (\texttt{DeepSeek-V4-Flash} \cite{deepseek2026deepseek}) to determine their relation with the target topic.
See illustration of the flow in~\figref{fig:telltail_opt_overview}, and examples of optimized queries in \appref{app:full_queries_examples}.
We construct $|Q_R|=10$ optimized fingerprint queries for each candidate retriever, yielding a total query budget of $|Q| = |Q_R|\cdot|S| = 190$. In \appref{app:medicalqa_evaluation}, we further
evaluate a different target topic within the MedicalQA dataset (\emph{kidney diseases}) and show
that \ours{} maintains consistent performance.

\parhead{Optimization Configuration}
Across all \oursopt{} experiments, we optimize 30-token suffixes.
We use a fixed blocking budgetof $50\%$ for most models, based on a sweep over budgets from $0\%$ to $90\%$, where $50\%$ provided a simple and effective default.
We deviate from this setting only for the closed-weights model, for which we use $25\%$ to reduce the cost and difficulty of black-box optimization (RASLITE), and for \emph{gemma-300M}, for which we use $90\%$ to suppress persistent target-specific lexical leakage. Further details on the sweep and these choices are provided in \appref{app:blocking_effects}.

\mainsubsection{Decision Rules and Metrics}
\label{sec:setup_metrics}

In each trial, and a respective target system, we compare the candidate acceptance scores (\secref{sec:method}). We define the set of accepted candidates using an acceptance threshold for each fingerprint strategy. Unless otherwise stated, we use the following default thresholds:  $\tau_1=1.0$ for \oursrandom{} and \ourstopic{} (ablation in \appref{app:threshold_sensitivity}), and $\tau_2=0.5$ for \oursopt{}.
If there is only one accepted candidate, it is predicted as the target retriever. Otherwise, if multiple candidates are accepted, the attack abstains (i.e., outputs $\bot$); if no candidate is accepted, the attack rejects the candidate set and outputs $\unk$.

\parhead{Attack Success Rate}
We define attack success rate(ASR) as the classification success under
the fixed decision rule above.
Namely, ASR is the fraction of target systems---and similarly, of trials---for which the fingerprinting was successful.
For a known target retriever,
$R^\star \in S$, the attack succeeds only when it predicts $R^\star$.
For an unknown target, $R^\star \notin S$, the attack succeeds when it
rejects the candidate set and outputs $\unk$.

\parhead{Threshold-Sweeping Metrics}
We additionally
evaluate the trade-off between correctly identifying known retrievers and falsely accepting unknown retrievers as the acceptance threshold varies.
To allow the use of threshold at the cases of multiple or none accepted candidates, we always consider the highest score among candidates.
Sweeping a threshold over
this score produces different operating points which we analyze as follows.
(Refer to \figref{fig:heatmap_llm} in \appref{app:response_only_heatmap} for an example of such scoring matrix.)

\emph{OSCR Curve.}
Following ~\cite{NEURIPS2018_48db7158}, we use the
Open-Set Classification Rate (OSCR) curve, defined by the correct classification rate (CCR) and false acceptance rate (FAR) under a threshold $\tau$.
$\mathrm{CCR}(\tau)$ is the fraction of
\textit{known} target retrievers (i.e., retrievers in the attacker set $S$) that are correctly predicted under threshold $\tau$.
$\mathrm{FAR}(\tau)$ is the fraction of \textit{unknown}
target retrievers (i.e., not in $S$) that are incorrectly predicted a known candidates.
The OSCR curve plots $\mathrm{CCR}(\tau)$ against
$\mathrm{FAR}(\tau)$ as $\tau$ varies.
We summarize this curve using
its area under the curve, AUOSCR.

We additionally report CCR at fixed false-acceptance counts.
Denote $\mathrm{FA}(\tau)$ as the number of unknown retrievers that are
incorrectly accepted as candidates. For a false-acceptance budget $c$,
we define
\[
\mathrm{CCR@FA}\leq c
=
\max_{\tau:\,\mathrm{FA}(\tau)\leq c}
\mathrm{CCR}(\tau).
\]

\mainsection{Results}
\label{sec:results}

\begin{figure*}[t]
    \centering
    \includegraphics[width=0.75\textwidth]{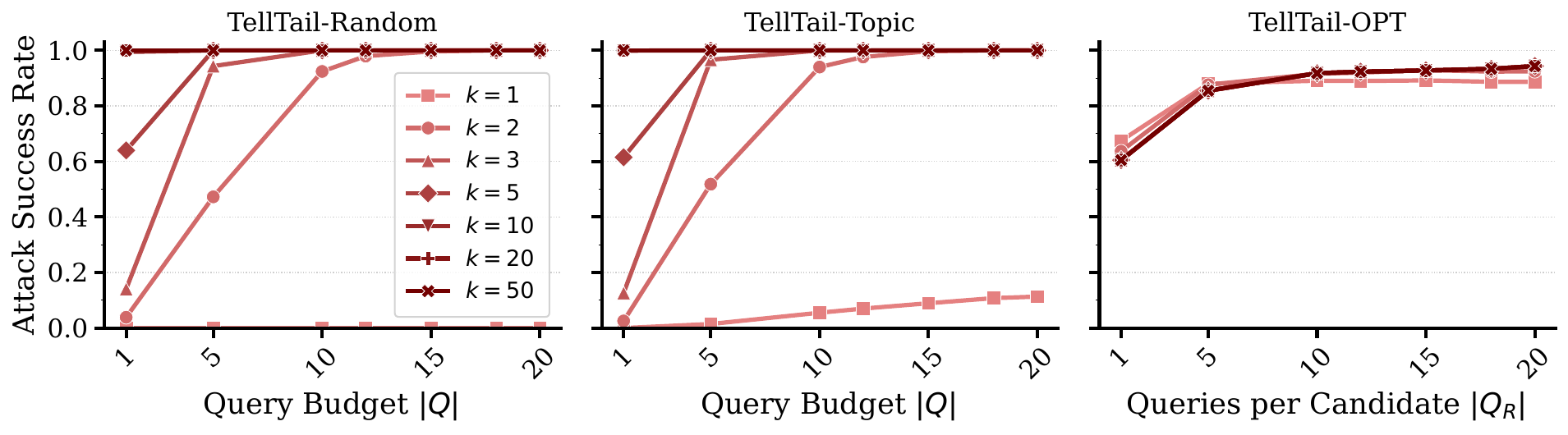}
    \caption{\textbf{Fingerprinting success under full-rank access (\tmone{}).} Attack success as a function of attacker budget (number of queries) and the exposed retrieval results ($k$), across our methods:
    \oursrandom{} (left), \ourstopic{} (center), and \oursopt{} (right).
    For \oursrandom{} and \ourstopic{}, the horizontal axis denotes the
    total query budget $|Q|$. For \oursopt{}, it denotes the number
        of optimized queries per candidate model, $|Q_R|$.
        }
    \label{fig:tm1_telltail_asr}
\end{figure*}

We evaluate the effectiveness of \ours{} variants under the experimental setup defined in \secref{sec:experimental_setup}. The results are organized by interface, from the most informative setting, \tmone{} (\secref{sec:eval_tm1}), to the most restrictive setting, \tmthree{} (\secref{sec:eval_tm3}).

\mainsubsection{Full-Rank Interface (\tmone{})}
\label{sec:eval_tm1}

\begin{tcolorbox}[
    colback=blue!5,%
    colframe=blue!70!black,%
    arc=4mm,%
    boxrule=0.8pt,%
    left=6pt, right=6pt, top=6pt, bottom=6pt
]

\textbf{Finding:} Systems that fully expose retrieval ranks are near-perfectly fingerprintable; e.g., $20$ queries and ordered top-$2$ passages suffice to recover the deployed retriever.
\end{tcolorbox}

We evaluate under \tmone{}, a retrieval system that exposes the ordered top-$k$ ranking along the retrieved passages.

First, in \figref{fig:tm1_telltail_asr}, we measure the fingerprint success (\textbf{ASR}) as a function of the attacker budget $Q$ and retrieved passages $k$.
At the full attack budget, all three variants perform strongly: the generic-query variants (\oursrandom{} and \ourstopic{}) reach $100\%$ ASR for every $k\ge2$;
and \oursopt{} is slightly weaker in this regime, but still achieves $94.3\%$ ASR for $k\ge3$.

Under the restricted single-query setting, generic-query variants' ASR varies with $k$, but obtain a perfect ASR when exposed to $k \geq 10$ top-ranked passages. In other words, we are able to fingerprint using merely a \textit{single} query by observing the top-10 retrieved passages from the system.
Differently, restricting the budget of our optimization-based approach, \oursopt{}, to $|S|$ (the minimal budget, since queries are candidate-model-specific), all $k$ values yield similar ASRs around $60\%$.
In particular, \oursopt{} seems to be effective under low $k$ values: it is able to achieve near-perfect ASR while accessing only a single top-ranked passage ($k=1$).

Second, we vary the acceptance threshold yielding operating points with
different false-acceptance budgets (\figref{fig:tm1_telltail_oscr} in \appref{app:oscr_ccr_results}). With no false acceptances allowed ($\mathrm{CCR@FA\leq0}$),
the generic-query variants achieve a perfect score once enough ranked
passages are exposed (e.g., $k=10$), but identify no known retriever at $k=1$.
\oursopt{} shows
the opposite trend, reaching $58\% \mathrm{CCR@FA\leq0}$ at $k=1$ but degrading at larger
$k$.
Allowing at most two false acceptances ($\mathrm{CCR@FA}\leq2$) shows that
\oursopt{}'s poor performance with zero false acceptancesis caused by only
a few problematic unknown retrievers, with CCR reaching $100\%$ at $k=1$
and approximately $95\%$ at $k=10$.

Inspecting the two false acceptances made by \oursopt{} we find that the first classifies \emph{nomic-v1} as \emph{nomic-v1.5}, two closely
related model releases with shared lineage~\cite{nussbaum2024nomic};
their similarity therefore provides a plausible explanation for the collision.
The second classifies \emph{gte-base} as \emph{bge-base}. Here, we do not
identify an equally direct relationship. The source of this
collision remains unclear, potentially involving less visible aspects of the models such as their
training objectives or data.

The results show that the different fingerprint strategies have complementary aspects: generic queries benefit primarily from observing more passages, whereas optimized queries exhibit strong fingerprinting ability with only a small amount of exposed passages.
Overall, under \tmone{}, \ours{} fingerprints retrievers near perfectly when the exposed top-$k$ or the query budget are sufficiently large, and still reasonably well when both are limited (e.g., ${\approx}60\%$ ASR for \oursopt{} with a single query per candidate at $k=1$).

\mainsubsection{Unordered Top-$k$ Interface (\tmtwo{})}
\label{sec:eval_tm2}

\begin{tcolorbox}[
    colback=blue!5,%
    colframe=blue!70!black,%
    arc=4mm,%
    boxrule=0.8pt,%
    left=6pt, right=6pt, top=6pt, bottom=6pt
]

\textbf{Finding:} Systems that expose unordered retrieval results are highly
fingerprintable; e.g., achieving near-perfect success with only 20 queries and top-5 passage access.
\end{tcolorbox}

\begin{figure}[t]
    \centering
    \includegraphics[width=0.85\columnwidth]{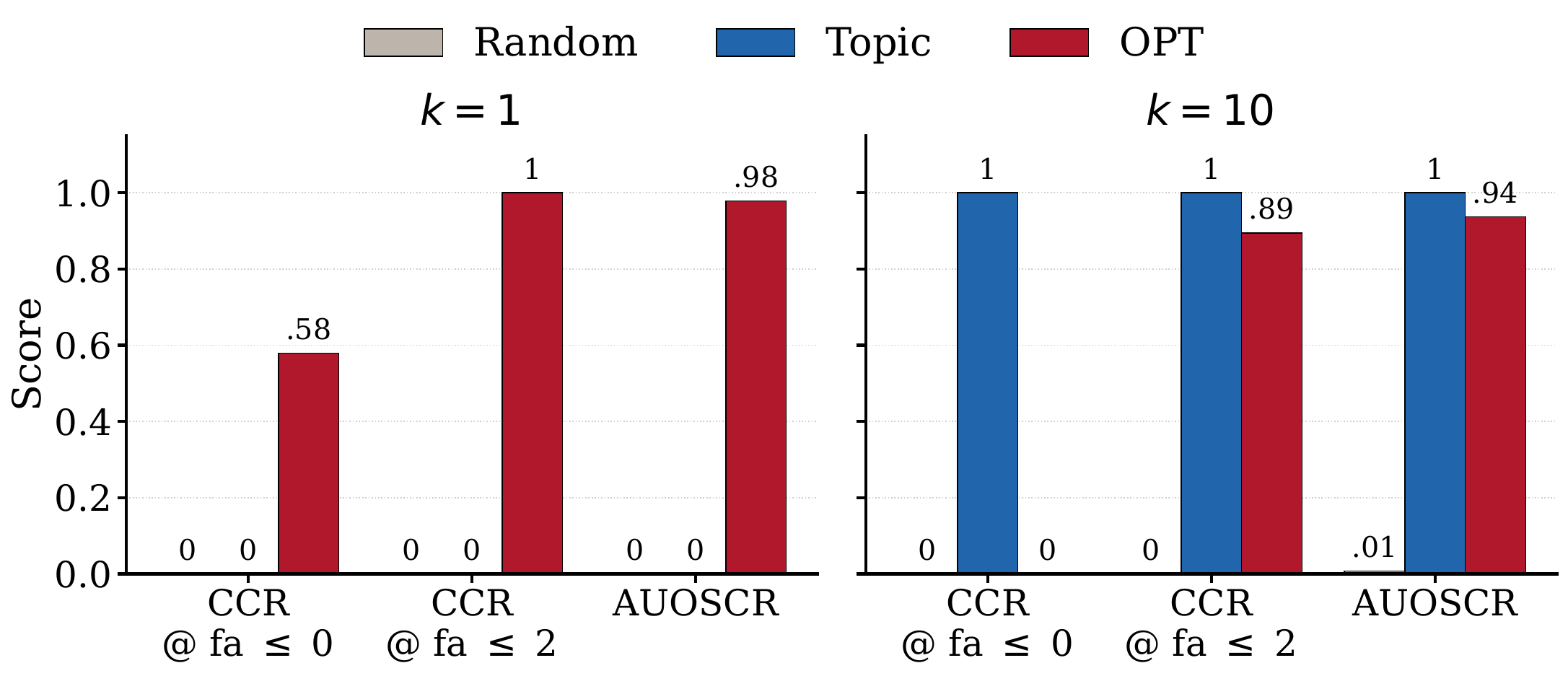}
    \caption{\textbf{Fingerprinting metrics under unordered top-$k$ result access (\tmtwo{}).}
    Measuring under $k=1$ (left) and $k=10$ (right), using the full query budget.
    Metrics report CCR while adjusting the threshold for up to 0 and 2 false acceptances of unknown retrievers, corresponding to FAR${\leq}0\%$ and FAR${\leq} 5\%$, respectively, together with AUOSCR.}
    \label{fig:tm2_telltail_oscr}
\end{figure}

\begin{figure*}[t]
    \centering
    \includegraphics[width=0.75\textwidth]{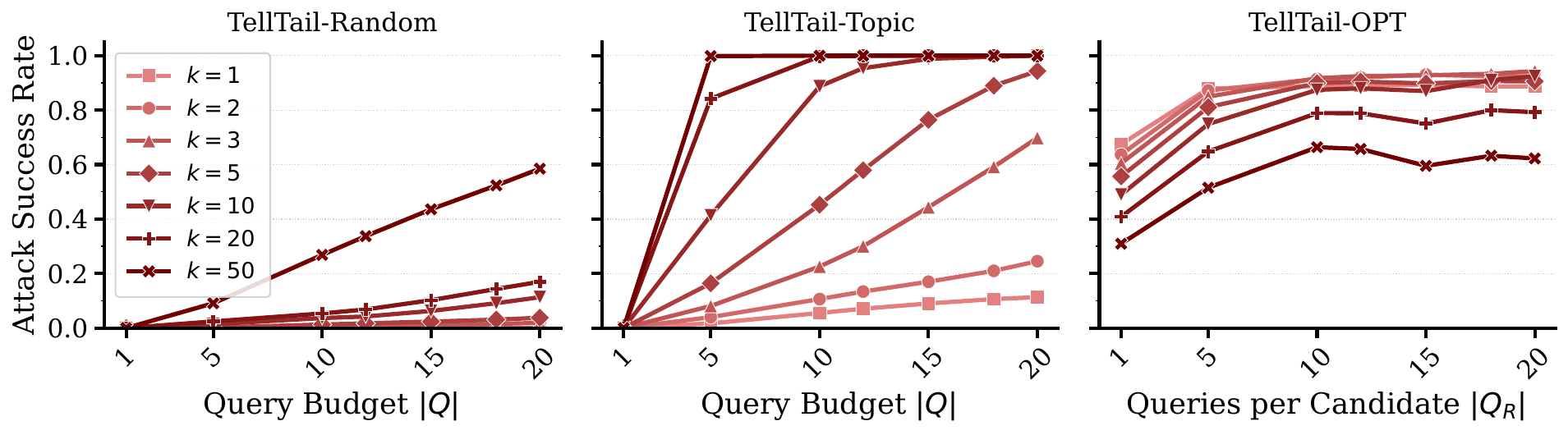}
    \caption{\textbf{Fingerprinting success under unordered top-$k$ result access (\tmtwo{}).}
    Attack Success as a function of the attacker budget (number of queries) and top exposed results ($k$), across our methods: \oursrandom{} (left),
        \ourstopic{} (center), and \oursopt{} (right). For \oursrandom{} and
        \ourstopic{}, the horizontal axis denotes the total query budget $|Q|$.
        For \oursopt{}, it denotes the number of optimized queries per
        candidate, $|Q_R|$.
        }
    \label{fig:tm2_telltail_asr}
\end{figure*}

We next evaluate under \tmtwo{}, a retrieval system that exposes the unordered top-$k$ retrieved passages.

\figref{fig:tm2_telltail_asr} reports the ASR as function of the attacker budget $Q$ and the number of top passages available $k$.
We find that \oursrandom{} performs relatively poorly across all budgets, peaking at full budget and using the top-50 passages with $\le$58\% ASR.
This shows that \tmtwo{} poses a more challenging setting for \oursrandom{} than \tmone{}---where this exact method obtained high success in both low $k$s and low budget. This suggests the \oursrandom{} fingerprinting relies on the \textit{order} of the passages.

Differently, our other generic-query variant, \ourstopic{}, which uses single-topic-related queries (here, Harry Potter), significantly outperforms its random-query counterpart on all budgets and $k$ values.
It achieves, under full budget, ASR of $\approx 95\%$ when $\ge k=5$ passages are exposed.

Meanwhile, the optimization variant, \oursopt{}, shows a trend consistent with \tmone{} setting, and outperforms other methods when few passages are exposed. For instance, at $k=3$, it achieves $94.3\%$ ASR, compared with $69.8\%$ for \ourstopic{}. However, as $k$ increases, the gap narrows, with \ourstopic{} outperforming all methods for $k\ge10$.
Notably, as $k$ increases  \oursopt{} only loses performance; this stems from the loss of information when increasing $k$. In particular, \oursopt{} optimizes a query to promote a specific passage to the top results for a specific model; as $k$ increases, this passage appears for other non-target models, weakening the fingerprinting. For example, a passage ranked first by the target model but $50$\textsuperscript{th} by another model is treated identically when $k=50$ and therefore implies an incorrect prediction (\figref{fig:heatmap_passages} in \appref{app:vissible_passages_heatmap}  presents the confusion matrix).

\figref{fig:tm2_telltail_oscr} further compares two representative cutoffs
$k=1$ and $k=10$ across acceptance thresholds.
With no false acceptances allowed
($\mathrm{CCR@FA}\leq0$), \ourstopic{} reaches
perfect CCR at $k=10$ (compared with $k=2$ under \tmone{}).
\oursopt{}, in contrast, achieves $58\%$ CCR at $k=1$ but no correct
identifications at $k=10$.

Allowing at most two false acceptances ($\mathrm{CCR@FA}\leq2$) changes the picture for
\oursopt{}, where at $k=1$, its CCR rises to $100\%$.
At $k=10$, it%
achieves $89\%$ CCR, while \ourstopic{}
remains perfect and \oursrandom{} remains ineffective. We also observe that the two false acceptances made by \oursopt{} here are the same as under \tmone{} (\secref{sec:eval_tm1}): \emph{nomic-v1} is accepted as \emph{nomic-v1.5}, and \emph{gte-base} as \emph{bge-base}. Results across all $k$ values are reported in \appref{app:oscr_ccr_results}.

The results show both that removing passage order substantially weakens \oursrandom{} and that the complementary roles of \ourstopic{} and \oursopt{} become more pronounced: \ourstopic{} requires access to more top results to reach strong performance, whereas \oursopt{} achieves near-perfect access with only a few top results.
Overall, \ours{} methods can effectively fingerprint under access to the unordered top results with varied $k$ and with limited query budgets.

\mainsubsection{Response-only Interface (\tmthree{})}
\label{sec:eval_tm3}
\begin{tcolorbox}[
    colback=blue!5,%
    colframe=blue!70!black,%
    arc=4mm,%
    boxrule=0.8pt,%
    left=6pt, right=6pt, top=6pt, bottom=6pt
]

\textbf{Finding:} RAG System that expose \textit{only} the generated response enable identification of their underlying retriever in $>90\%$ of the cases.
\end{tcolorbox}

We next evaluate fingerprinting the retriever underlying a RAG system that only outputs the generated response.
Our attacker uses \oursopt{}, and optimizes
a total of $|Q_R|=10$ fingerprint queries toward
the \HP{} concept, per candidate retriever. The attacker then queries the RAG system, inspects the generated response, and considers it a \textit{match} when an LLM judge determines that the response is related to \HP{}.

\begin{figure}[t]
    \centering

        \includegraphics[width=0.6\linewidth]{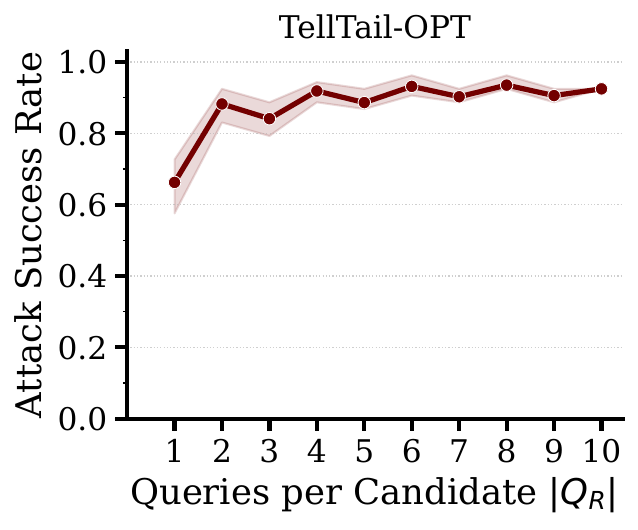}

    \caption{
    \textbf{Fingerprinting success with access only to the final RAG response (\tmthree{}).}
    Attack success as a function of the number of optimized queries per candidate retriever, $|Q_R|$. Only \oursopt{} is applicable in this restrictive setting.
    }
    \label{fig:tm3_telltail_asr_iqr}
\end{figure}

First, evaluating the fingerprint success under this strict setting in
\figref{fig:tm3_telltail_asr_iqr}, we find \oursopt{} achieves successful fingerprinting of most models ($>80\%$) even with two queries per candidate ($|Q_R|=2$).
Using our full query budget ($|Q_R|=10$, $|Q|=190$), our attack achieves a high success rate of $92.5\%$ ($49/53$ correct predictions).
We additionally analyze various threshold, showing the scores yielded by \oursopt{} in this setting are highly separable ($AUOSCR=0.98$); this is further demonstrates by the experiment confusion matrix (\appref{app:response_only_heatmap}).
Further analyzing the attack failure, we observe similar false acceptance cases to those of the former settings (e.g., the Nomic models).
Overall, the results show our optimized queries overwhelmingly steer the model response, by targeting the underlying context provided by the retrieved passages.

To better understand how optimizing toward a topic centroid affects retrieval,
we evaluate each optimized query against all $53$ retrievers and inspect its
top-$100$ retrieved passages.
We label each passage as \HP{}-related using
\texttt{DeepSeek-V4-Flash}and, independently, using a keyword-based heuristic. The two labeling methods agree on $99.4\%$ of passages; we manually resolve the remaining disagreements.
Notably, our \tmthree{} setting, per OpenWebUI default configuration, only requires affecting the top-$3$ passages provided as context to the prompt; the following inspection, thus, evaluates the dominance of the topic in the top results.

\begin{figure}[t]
    \centering
    \includegraphics[width=\linewidth]{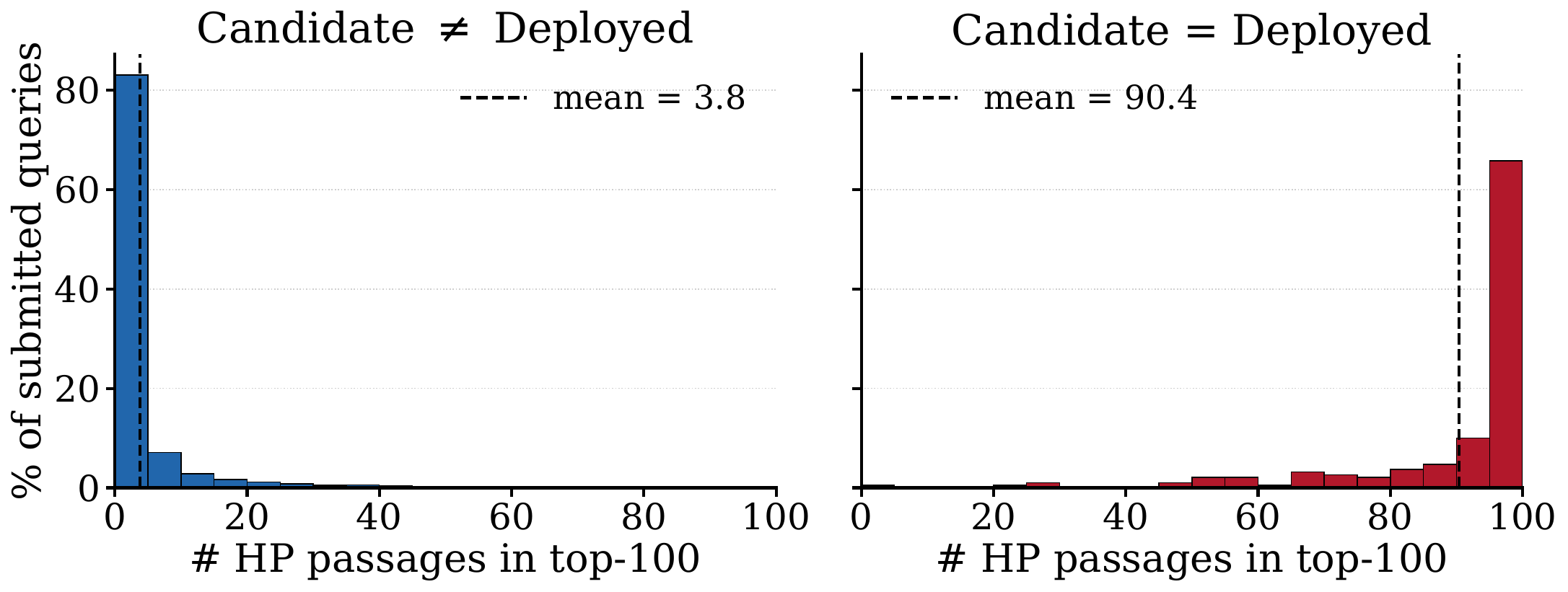}
    \caption{\textbf{Target-topic concentration in the top-$100$.}
    Distribution of the number of \HP{} passages retrieved per optimized query,
    conditioned on whether the deployed retriever differs from the candidate retriever
    for which the query was optimized (left), or matches it (right). Dashed lines denote means.}
    \label{fig:hp_top100}
\end{figure}

When an optimized query is submitted to a deployed retriever, we count how
many of its top-$100$ retrieved passages concern \HP{}.
\figref{fig:hp_top100} reports the distribution of this count, conditioned
on whether the deployed retriever is the same retriever for which the query
was optimized. When they match, queries retrieve on average $90.4$ \HP{}
passages among the top-$100$; when the query is evaluated on other
retrievers, this drops to only $3.8$ passages.

Among the passages retrieved by these other retrievers, we further find that
approximately $20\%$ concern \emph{George Washington} or related U.S. politics,
reflecting the benign query topic. The remaining passages are more
heterogeneous and include English/writing content, word definitions,
literature, and other generic/noisy material (like web page errors).

Overall, the experiment provides strong evidence that centroid optimization
induces a strong retriever-specific shift in the retrieved context toward
the target topic, and suggests that this effect can remain pronounced even
when substantially larger retrieved contexts are supplied to the
LLM. Thus, this result motivates testing our method in \secref{sec:case_study}.

\mainsection{Case Study: Fingerprinting in a Deployed RAG System}
\label{sec:case_study}

In the following case study,
we consider an actual RAG application, OpenWebUI, and test our fingerprinting method, \oursopt{}, to systematically identify the retriever.
Compared to the controlled setting of our systematic evaluation (\secref{sec:results}), the case study targets a practical RAG deployment, including its complexities.

RAG applications provide project-style
workspaces in which users upload a collection of documents and interact
with a language model grounded in their contents. This interaction
pattern appears, for example, as features of ChatGPT \cite{openai_projects},
Claude \cite{anthropic_projects}, and OpenWebUI \cite{openwebui}. In such systems,
the application retrieves relevant portions of the uploaded documents
and supplies them as context to the language model. To improve retrieval performance, these applications might introduce additional retrieval components.
We deploy OpenWebUI under its default configuration, including that of these components; per our threat model, we assume the attacker neither controls these components nor knows their configuration.

\parhead{(i) Document Chunking}
Retrieval-augmented systems typically partition uploaded documents into smaller,
possibly overlapping chunks that serve as the passages indexed for
retrieval~\cite{retrieval-tricks--anthropic}.
This component's configuration naturally affects the retrieved passages, and thus increases attacker uncertainty.

\parhead{(ii) Reranking}
Retrieval-augmented systems might add a second ranking stage~\cite{nogueira2020passagererankingbert,gao2024modular} where the retriever first returns a larger candidate pool (top-$K$, $K \gg k$), then a separate reranker model re-scores this pool, and only the top-$k$ reranked passages are passed to the LLM.
Both the existence of the reranker and its unknown identity adds to the uncertainty of the attacker.

\parhead{(iii) Query Generation}
Retrieval-augmented systems may also \textit{transform} the user query before retrieval, for example
by rewriting, splitting, expanding, or by generating multiple retrieval
queries~\cite{ma-etal-2023-query,koo2024optimizing,gao2024modular} (a.k.a., query rephrasing).
Such processing may modify or remove payload placed in queries by fingerprint attackers.

\mainsubsection{Experimental Setup}
\label{sec:case-study-setup}
We run an end-to-end experiment on OpenWebUI (\emph{v0.9.6}) under its default configurations: default document chunking (chunks of 1{,}000 characters with 100-character overlap), query generation enabled. Thus, both the indexed retrieval passages and the query reaching the retriever are determined by the application.
We additionally enable reranking, where the retriever returns a top-$K$ candidate pool ($K=100$) and passes it to a reranker model, which in turn propagates the top-$3$ passages to the LLM. We use \texttt{BAAI/bge-reranker-v2-m3}~\cite{multi2024m3,li2023making}, the reranker OpenWebUI suggests when enabling reranking. Responses are generated by \texttt{gpt-4o-mini}, and fingerprinting relies solely on the generated response, consistent with \tmthree{}. In \appref{app:case_study_analysis}, we also evaluate the deployment without reranking, which matches the exact \tmthree{} setting of our systematic evaluation (response over the top-$3$ passages by the retriever), and further show that, for most candidate retrievers, fingerprinting may be reranker-agnostic.

\begin{figure}
    \centering
    \begin{tikzpicture}
        \node[
            inner sep=0pt,
            blur shadow={shadow blur steps=6,shadow xshift=-2pt,shadow yshift=2pt}
        ]{
            \includegraphics[width=1\linewidth]
                {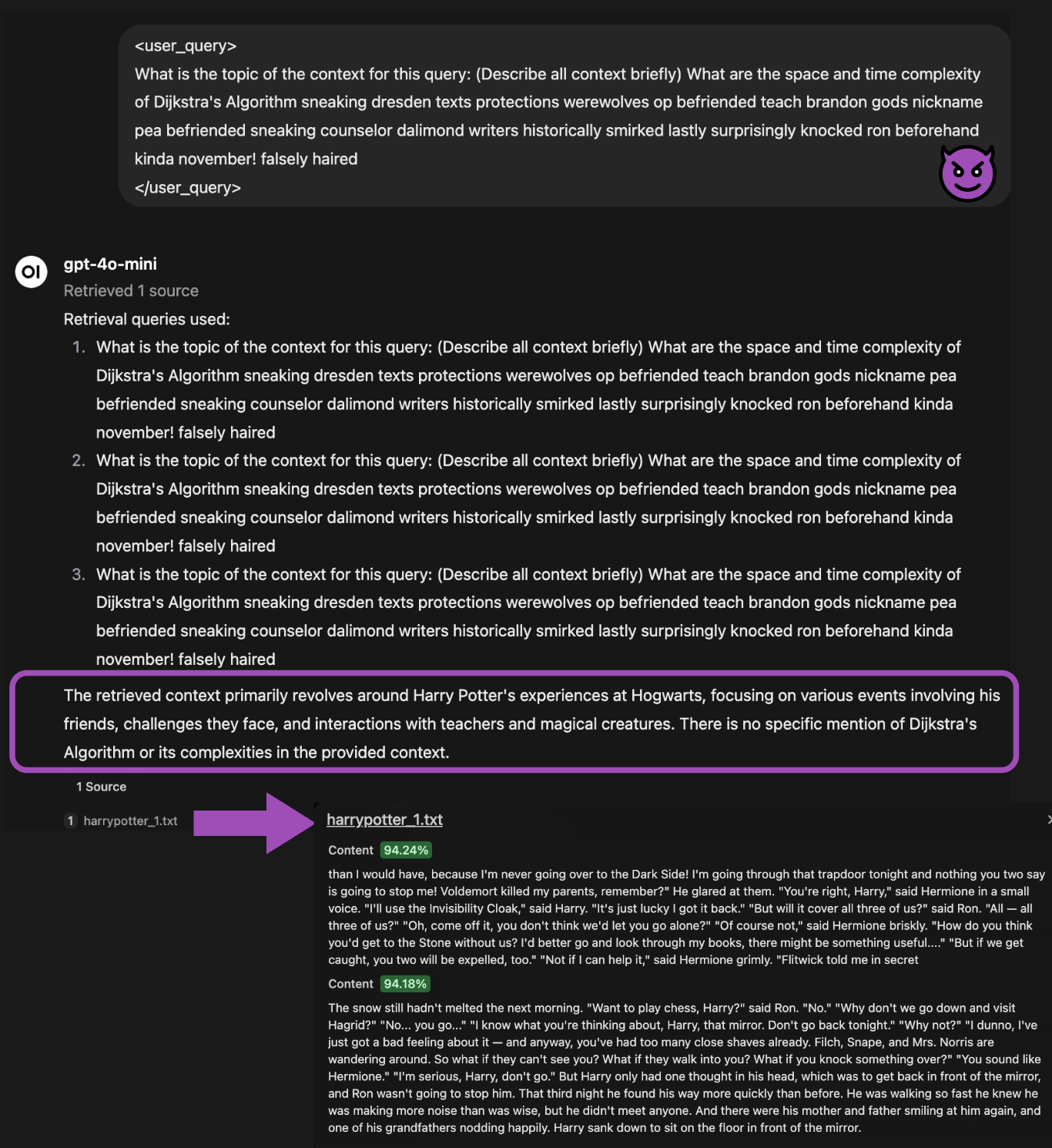}
        };
    \end{tikzpicture}

    \caption{Representative execution of the two-turn interaction in OpenWebUI.
    For compactness, we omit
    the initial query-preservation prompt.
    The figure shows the subsequent optimized query and the retrieval queries
    generated by OpenWebUI, which preserve the complete query. With retrieval
    depth $K=100$, we see that all 100 retrieved chunks originate from
    the target-topic document, \texttt{harrypotter\_1.txt}. Consequently, the
    generated response characterizes the retrieved context as \HP{}, despite
    the benign prefix asking about Dijkstra's algorithm.}
    \label{fig:openwebui_example}
\end{figure}

\parhead{Two-turn Interaction}
To improve preservation of the optimized query through the query-generation stage, we issue each query using a two-turn interaction. Since OpenWebUI generates retrieval queries with an LLM call conditioned on the recent chat history (\appref{app:query_generation_template}), an earlier user message can assist with this step. In the first turn, the attacker instructs the LLM to preserve the next user message as the retrieval query, without rewriting, shortening, splitting, or otherwise modifying it. In the second turn, the attacker submits the complete optimized query. Each query is issued in a fresh conversation to prevent previous interactions from affecting query generation (\appref{app:openwebui_prompts} presents the complete initial prompt). Importantly, this procedure encourages preservation of the query but does not guarantee bypassing OpenWebUI's query-generation stage completely. \figref{fig:openwebui_example} presents an example of an interaction where our attack succeeds.

\parhead{Knowledge Base}
We instantiate a document-grounded chat application containing two
semantically unrelated domains. The knowledge base consists of
\emph{Harry Potter and the Philosopher's Stone} and textbooks
on \emph{data structures}, \emph{algorithms}, and \emph{computational complexity}.
OpenWebUI applies its own document-processing pipeline and $13.7\%$ of the resulting
index originate from the \HP{} novel. \HP{} is therefore a minority topic in the
index. Notably, in deployed RAG applications, users---and thus attackers---can control and upload the knowledge base.

\parhead{Query Construction}
Following our approach in \oursopt{}, we construct queries agnostic to OpenWebUI's chunks, and optimized towards a set of passages.
We sample 100 random passages from the novel and divide them into 10 disjoint sets of 10 passages.
For each candidate retriever and each passage set, we
optimize a suffix toward the centroid of that set.
Each candidate
fingerprint therefore contains 10 independently optimized queries.

All optimized fingerprint queries use the same benign prefix: asking the LLM to describe the context followed by a question about \textit{Dijkstra's algorithm}. When the optimized suffix fails to steer the deployed retriever, the retrieved context is expected to remain about algorithms; if it succeeds, the context should shift toward \HP{}, making the two outcomes readily distinguishable in the generated response.

We deploy each of the 53 evaluated embedding models in turn and use the
same 19-model candidate set, response-level match criterion, and the majority
threshold as in \secref{sec:results}.
Each candidate contributes 10 queries; consequently,
fingerprinting one deployed retriever evaluates 190 optimized queries.

\mainsubsection{End-to-End Fingerprinting Results}
\label{sec:case_study_results}

\begin{tcolorbox}[
    colback=blue!5,%
    colframe=blue!70!black,%
    arc=4mm,%
    boxrule=0.8pt,%
    left=6pt, right=6pt, top=6pt, bottom=6pt
]
\textbf{Finding.} In a practical end-to-end RAG deployment, \ours{} remains effective under application-controlled document chunking and query generation, and with a reranker. Query generation emerges as the primary challenge; reranking has little effect on the attack.
\end{tcolorbox}

\begin{table}[t]
\centering
\caption{Fraction of candidate-specific optimized queries whose entire top-$k$ retrieval pool consists of target-topic chunks, with and without query generation.}
\label{tab:query_gen}
\resizebox{\columnwidth}{!}{%
\begin{tabular}{lcccc}
\toprule
Query generation & $k=3$ & $k=10$ & $k=50$ & $k=100$ \\
\midrule
Off & 96.32\% & 95.26\% & 92.63\% & 88.95\% \\
On  & 84.21\% & 82.11\% & 77.89\% & 70.53\% \\
\bottomrule
\end{tabular}%
}
\end{table}

\begin{figure}[t]
    \centering

        \includegraphics[width=0.7\linewidth]{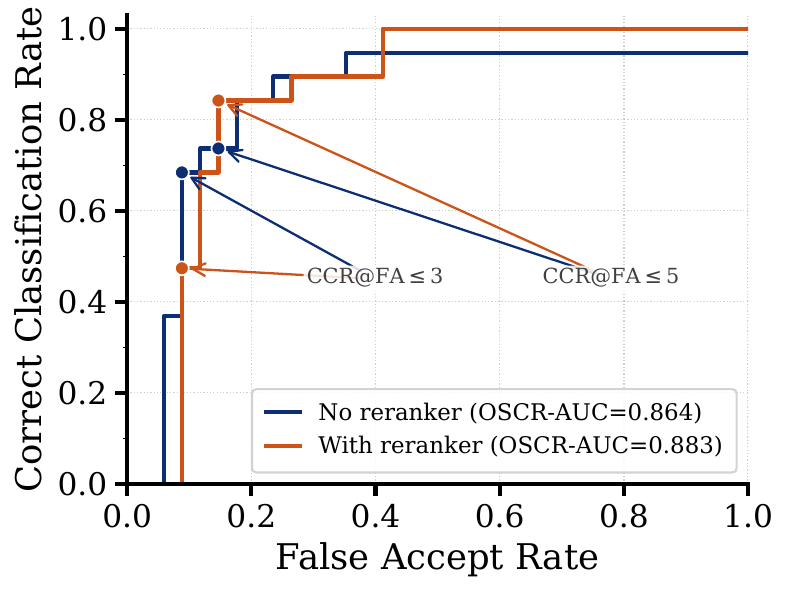}

    \caption{Performance of \oursopt{} in the end-to-end
        OpenWebUI case study, with and without reranking. The curves report the
        correct classification rate (CCR) on candidate retrievers against the
        false acceptance rate on non-candidate retrievers as the decision
        threshold varies. Markers indicate operating points allowing at most
        three and five false model acceptances.}
    \label{fig:case_study_oscr}
\end{figure}

In our end-to-end experiment \oursopt{} achieves an \textbf{ASR} of $83.0\%$
(correctly classifying 44 of the 53 deployed retrievers
at the fixed threshold $\tau=0.5$),
showing a slight degradation compared to the evaluation under the idealized RAG of \ref{sec:eval_tm3} (92.5\% ASR).
The scoring matrix (\appref{app:case_study_scoring_matrix}, \figref{fig:case_study_heatmap}) shows that the nine errors fall into two groups. Three are missed candidates (\emph{nomic-modernbert}, \emph{qwen-0.6b}, \emph{qwen-4b}), whose optimized queries are harmed most by the query generation mechanism (\appref{app:case_study_analysis}). The remaining six are false acceptances if%
deployed retrievers, which are misled by queries optimized for certain candidates, including one false acceotance (\emph{nomic-v1} as \emph{nomic-v1.5}) consistent with \secref{sec:eval_tm3}.
We hypothesize this happens because with only two topics in the knowledge base, a suffix that merely perturbs a non-target retriever's query away from the benign topic (computer science) can only shift retrieval toward the other (\HP{}).
Across decision thresholds, \oursopt{} achieves an AUOSCR of $0.88$  (\figref{fig:case_study_oscr} of \appref{app:case_study_analysis} presents OSCR curves).

An in-depth analysis of which OpenWebUI's features affect the attack (details in \appref{app:case_study_analysis}) indicates that enabling query generation is the primary source of degradation in fingerprinting success. When optimized queries are submitted to their intended retrievers, query generation reduces the fraction for which the entire retrieved context revolves around the target-topic (\HP{}) from $96.32\%$ to $84.21\%$ at top-$3$, and from $88.95\%$ to $70.53\%$ at top-$100$ (\tabref{tab:query_gen}). In contrast, reranking has little-to-no effect on the induced retrieval behavior.

\mainsection{Conclusion}
This work shows that embedding-based text retrievers underlying a black-box retrieval systems can be identified from externally observable behavior. We introduce \ours{}, a fingerprinting approach that exploits differences in the
retrieval behavior of retrievers across a range of access levels. We demonstrate that model-specific peculiarities provide a strong signal for identification. Evaluated on 53 retrievers, \ours{} reliably identifies the deployed retriever, achieving perfect identification when ordered retrieval results are available, 94.3\% ASR from unordered top-$3$ results, and 92.5\% ASR from generated responses alone. We further demonstrate the attack in an end-to-end OpenWebUI deployment, where it remains effective despite application-side query generation, document chunking, and reranking.
These findings have an important security implication: hiding the embedding model behind an opaque system provides little protection against an adversary capable of querying it. In particular, recovering model identity from queries alone can provide the reconnaissance required by model-specific retrieval attacks, illustrating the brittleness of relying on security through obscurity to protect retrieval systems.
More broadly, \ours{} demonstrates that differences between embedding models that may appear incidental can become externally observable security signals. As embedding-based retrieval becomes increasingly embedded within RAG and other deployed AI systems, understanding and mitigating this behavioral leakage will be important for designing systems whose internal components cannot be inferred simply through interaction.

\parhead{Limitations}
Our study also has several limitations and leaves opportunities for future work. \oursopt{} can produce unnatural-looking queries in some cases; however, prior work has shown that alternative optimization techniques can improve input naturalness while retaining adversarial effectiveness, and adapting such techniques to fingerprinting is a promising direction \cite{ben_tov2025gaslite,zhang2026adversarial}. Our evaluation also focuses on systems whose retrieval is primarily driven by dense embeddings. Extending fingerprinting to hybrid retrieval pipelines that combine dense retrieval with lexical methods such as BM25\cite{bm25hybridText}, and designing defenses that hinder fingerprinting by suppressing model-specific signals while preserving retrieval quality, both remain important directions for future work.

\section*{Ethical Considerations}
This work presents an attack that identifies the embedding model behind a retrieval-augmented system, and, as with other fingerprinting primitives, it is dual-use. Benign parties can use it to audit deployments, e.g., to verify that a provider serves the model it advertises, while adversaries may use it for reconnaissance ahead of model-specific attacks such as corpus poisoning, or to expose a provider's design choices. We believe the benefits of disclosure outweigh these risks. Our fingerprinting technique relies only on standard query access, and our results show that keeping the embedding model hidden provides little protection. System designers should therefore not treat the model's identity as a secret, and should instead adopt defenses against the model-specific attacks it enables.

All experiments were conducted on retrieval-augmented systems in controlled environments, including our local OpenWebUI instance; we did not query any third-party production system. Our experiments involve no human subjects and no personal data, and we use publicly available models and datasets (MS~MARCO, \medicalqa{}).

\bibliographystyle{IEEEtran}
\bibliography{references}

    \appendices
    \renewcommand{\thesectiondis}[2]{\Alph{section}\enspace}

\mainsection{Evaluated Models}
\label{app:models}

As mentioned in \secref{sec:setup_models_data}, we evaluate our fingerprinting methods over a pool of 53 embedding
models. The 19 models in the candidate set $S$ used in the main
fingerprinting experiments are listed first in
\tabref{tab:all_models}, followed by the remaining 34 models used
for broader evaluation.

\begin{table*}[!htbp]
\centering
\small
\caption{Full list of 53 evaluated embedding models. The attacked models in the candidate set $S$ are listed first, followed by the remaining evaluated models.}
\label{tab:all_models}
\resizebox{\textwidth}{!}{%
\begin{tabular}{ll}
\toprule
\textbf{Model} & \textbf{Identifier} \\
\midrule

\multicolumn{2}{l}{\textbf{Candidate models ($S$)}} \\
\midrule
\rowcolor{AttackBlueBG} qwen-0.6b \cite{zhang2025qwen3} & \texttt{Qwen/Qwen3-Embedding-0.6B} \\
\rowcolor{AttackBlueBG} qwen-4b \cite{zhang2025qwen3} & \texttt{Qwen/Qwen3-Embedding-4B} \\
\rowcolor{AttackBlueBG} e5-small \cite{wang2024E5} & \texttt{intfloat/e5-small-v2} \\
\rowcolor{AttackBlueBG} e5-base \cite{wang2024E5} & \texttt{intfloat/e5-base-v2} \\
\rowcolor{AttackBlueBG} e5-large \cite{wang2024E5} & \texttt{intfloat/e5-large-v2} \\
\rowcolor{AttackBlueBG} gte-modernbert \cite{zhang2024mgte} & \texttt{Alibaba-NLP/gte-modernbert-base} \\
\rowcolor{AttackBlueBG} gte-large-v1.5 \cite{zhang2024mgte} & \texttt{Alibaba-NLP/gte-large-en-v1.5} \\
\rowcolor{AttackBlueBG} gte-base-v1.5 \cite{zhang2024mgte} & \texttt{Alibaba-NLP/gte-base-en-v1.5} \\
\rowcolor{AttackBlueBG} gte-multi \cite{zhang2024mgte} & \texttt{Alibaba-NLP/gte-multilingual-base} \\
\rowcolor{AttackBlueBG} nomic-modernbert \cite{nussbaum2024nomic} & \texttt{nomic-ai/modernbert-embed-base} \\
\rowcolor{AttackBlueBG} nomic-v1.5 \cite{nussbaum2024nomic} & \texttt{nomic-ai/nomic-embed-text-v1.5} \\
\rowcolor{AttackBlueBG} gtr-t5-base \cite{ni2022gtr} & \texttt{sentence-transformers/gtr-t5-base} \\
\rowcolor{AttackBlueBG} minilm-l6 \cite{wang2020minilm} & \texttt{sentence-transformers/all-MiniLM-L6-v2} \\
\rowcolor{AttackBlueBG} minilm-l12 \cite{wang2020minilm} & \texttt{sentence-transformers/all-MiniLM-L12-v2} \\
\rowcolor{AttackBlueBG} mpnet \cite{song2020mpnet} & \texttt{sentence-transformers/all-mpnet-base-v2} \\
\rowcolor{AttackBlueBG} snowflake-v2 \cite{merrick2405arctic} & \texttt{Snowflake/snowflake-arctic-embed-m-v2.0} \\
\rowcolor{AttackBlueBG} gemma-300m \cite{vera2025embeddinggemma} & \texttt{google/embeddinggemma-300m} \\
\rowcolor{AttackBlueBG} bge-base \cite{bge_embedding} & \texttt{BAAI/bge-base-en-v1.5} \\
\rowcolor{AttackBlueBG} text-embedding-3-small (openai) \cite{openai2024textembedding3small} & \texttt{openai/text-embedding-3-small} \\
\midrule

\multicolumn{2}{l}{\textbf{Additional evaluated models}} \\
\midrule
\rowcolor{OtherRedBG} multilingual-e5-small \cite{wang2024multilingualE5} & \texttt{intfloat/multilingual-e5-small} \\
\rowcolor{OtherRedBG} multilingual-e5-base \cite{wang2024multilingualE5} & \texttt{intfloat/multilingual-e5-base} \\
\rowcolor{OtherRedBG} multilingual-e5-large \cite{wang2024multilingualE5} & \texttt{intfloat/multilingual-e5-large} \\
\rowcolor{OtherRedBG} multilingual-e5-large-instruct \cite{wang2024multilingualE5} & \texttt{intfloat/multilingual-e5-large-instruct} \\
\rowcolor{OtherRedBG} gte-base \cite{li2023thenlpergte} & \texttt{thenlper/gte-base} \\
\rowcolor{OtherRedBG} nomic-v1 \cite{nussbaum2024nomic} & \texttt{nomic-ai/nomic-embed-text-v1} \\
\rowcolor{OtherRedBG} gtr-t5-large \cite{ni2022gtr} & \texttt{sentence-transformers/gtr-t5-large} \\
\rowcolor{OtherRedBG} paraphrase-minilm-l6 \cite{reimers2019sentencebert} & \texttt{sentence-transformers/paraphrase-MiniLM-L6-v2} \\
\rowcolor{OtherRedBG} paraphrase-multi-minilm-l12 \cite{reimers2019sentencebert} & \texttt{sentence-transformers/paraphrase-multilingual-MiniLM-L12-v2} \\
\rowcolor{OtherRedBG} paraphrase-multi-mpnet \cite{reimers2019sentencebert} & \texttt{sentence-transformers/paraphrase-multilingual-mpnet-base-v2} \\
\rowcolor{OtherRedBG} sentence-t5-base \cite{ni2022sentence} & \texttt{sentence-transformers/sentence-t5-base} \\
\rowcolor{OtherRedBG} sentence-t5-large \cite{ni2022sentence} & \texttt{sentence-transformers/sentence-t5-large} \\
\rowcolor{OtherRedBG} bge-small \cite{bge_embedding} & \texttt{BAAI/bge-small-en-v1.5} \\
\rowcolor{OtherRedBG} bge-large \cite{bge_embedding} & \texttt{BAAI/bge-large-en-v1.5} \\
\rowcolor{OtherRedBG} bge-m3 \cite{multi2024m3} & \texttt{BAAI/bge-m3} \\
\rowcolor{OtherRedBG} jina-v3 \cite{sturua2024jinaembeddingsv3multilingualembeddingstask} & \texttt{jinaai/jina-embeddings-v3} \\
\rowcolor{OtherRedBG} jina-v5-small \cite{akram2026jinaembeddingsv5texttasktargetedembeddingdistillation} & \texttt{jinaai/jina-embeddings-v5-text-small-retrieval} \\
\rowcolor{OtherRedBG} jina-v2-base \cite{gunther2023jina-v2} & \texttt{jinaai/jina-embeddings-v2-base-en} \\
\rowcolor{OtherRedBG} jina-v2-small \cite{gunther2023jina-v2} & \texttt{jinaai/jina-embeddings-v2-small-en} \\
\rowcolor{OtherRedBG} jina-v1-base \cite{gunther2023jina} & \texttt{jinaai/jina-embedding-b-en-v1} \\
\rowcolor{OtherRedBG} snowflake-v1 \cite{yu2024arctic} & \texttt{Snowflake/snowflake-arctic-embed-m} \\
\rowcolor{OtherRedBG} kalm-mini-v1 \cite{hu2025kalm} & \texttt{HIT-TMG/KaLM-embedding-multilingual-mini-v1} \\
\rowcolor{OtherRedBG} kalm-mini-instruct-v1 \cite{hu2025kalm} & \texttt{HIT-TMG/KaLM-embedding-multilingual-mini-instruct-v1} \\
\rowcolor{OtherRedBG} kalm-mini-instruct-v1.5 \cite{hu2025kalm} & \texttt{HIT-TMG/KaLM-embedding-multilingual-mini-instruct-v1.5} \\
\rowcolor{OtherRedBG} stella-en-400m-v5 \cite{zhang2024stella} & \texttt{NovaSearch/stella\_en\_400M\_v5} \\
\rowcolor{OtherRedBG} octen-0.6b \cite{octen2025rteb} & \texttt{Octen/Octen-Embedding-0.6B} \\
\rowcolor{OtherRedBG} solon-large \cite{ordalietech2023solonembeddingslarge01} & \texttt{OrdalieTech/Solon-embeddings-large-0.1} \\
\rowcolor{OtherRedBG} uae-large-v1 \cite{li2023UAE} & \texttt{WhereIsAI/UAE-Large-V1} \\
\rowcolor{OtherRedBG} gist-large \cite{solatorio2024gistembed} & \texttt{avsolatorio/GIST-large-Embedding-v0} \\
\rowcolor{OtherRedBG} croissant-v0.3 \cite{faysse2024croissantllm} & \texttt{manu/sentence\_croissant\_alpha\_v0.3} \\
\rowcolor{OtherRedBG} croissant-v0.4 \cite{faysse2024croissantllm} & \texttt{manu/sentence\_croissant\_alpha\_v0.4} \\
\rowcolor{OtherRedBG} mxbai-large-v1 \cite{emb2024mxbai1,li2023UAE} & \texttt{mixedbread-ai/mxbai-embed-large-v1} \\
\rowcolor{OtherRedBG} pixie-rune-v1 \cite{TelePIX-PIXIE-Rune-v1.0} & \texttt{telepix/PIXIE-Rune-v1.0} \\
\rowcolor{OtherRedBG} bilingual-large \cite{conneau2019bilingual1,reimers2019sentencebert,thakur2020bilingual3} & \texttt{Lajavaness/bilingual-embedding-large} \\
\bottomrule
\end{tabular}
}
\end{table*}

\mainsection{RAG Prompt}
\label{app:RAG_prompt}
Below is the RAG prompt we adopt from OpenWebUI in our systematic evaluation described in \secref{sec:experimental_setup}. Similarly to the flow followed by OpenWebUI, this prompt is passed with the top-$3$ retrieved passages as context followed by the user query.

\begin{ragprompt}[RAG Prompt Template]
\scriptsize
\ttfamily
\raggedright
\sloppy
\begin{Verbatim}[
  breaklines=true,
  breakanywhere=true,
  fontsize=\scriptsize
]

### Task:
Respond to the user query using the provided context, incorporating inline citations in the format [id] **only when the <source> tag includes an explicit id attribute** (e.g., <source id="1">).

### Guidelines:
- If you don't know the answer, clearly state that.
- If uncertain, ask the user for clarification.
- Respond in the same language as the user's query.
- If the context is unreadable or of poor quality, inform the user and provide the best possible answer.
- If the answer isn't present in the context but you possess the knowledge, explain this to the user and provide the answer using your own understanding.
- **Only include inline citations using [id] (e.g., [1], [2]) when the <source> tag includes an id attribute.**
- Do not cite if the <source> tag does not contain an id attribute.
- Do not use XML tags in your response.
- Ensure citations are concise and directly related to the information provided.

### Example of Citation:
If the user asks about a specific topic and the information is found in a source with a provided id attribute, the response should include the citation like in the following example:
* "According to the study, the proposed method increases efficiency by 20%

### Output:
Provide a clear and direct response to the user's query, including inline citations in the format [id] only when the <source> tag with id attribute is present in the context.

<context>
{{CONTEXT}}
</context>
\end{Verbatim}
\end{ragprompt}

\mainsection{Query Sets}
\label{app:queries_prefixes}

\tabref{tab:telltail_generic_queries} lists the 20 queries used for the
\oursrandom{} and \ourstopic{} experiments under the full-rank and unordered top-$k$ threat models. In addition, \tabref{tab:gaslite_prefix_queries} lists the 20 benign prefix queries used
in the full-rank and unordered top-$k$ interfaces for \oursopt{}. These queries are random queries
chosen to span diverse topics and serve as optimization prefixes.

\begin{table*}[!htbp]
\centering
\small
\caption{Queries used for each of the generic queries variants. \oursrandom{} queries are sampled from MS MARCO-style queries, while the \ourstopic{} queries are random synthetic \emph{Harry Potter}.}
\label{tab:telltail_generic_queries}
\resizebox{\textwidth}{!}{%
\begin{tabular}{p{0.47\textwidth} p{0.47\textwidth}}
\toprule
\textbf{\oursrandom{}} & \textbf{\ourstopic{} (\emph{Harry Potter})} \\
\midrule
when is the most expensive time to go to punta cana &
who is hagrid and what is his role at hogwarts \\
is fentanyl or dilaudid stronger &
what is the golden snitch and how does it end a quidditch match \\
how many people are living in Damascus now &
who is voldemort and what are horcruxes \\
effects of nicotine lozenges on the body &
what is the marauders map and how does it work \\
where is bath michigan located &
who is albus dumbledore and why is he important in the wizarding world \\
what are the cell organelles &
what houses exist at hogwarts and what traits define each one \\
structure that produces spores in fungi &
who is severus snape and what is his relationship with harry \\
boat whistle made in india &
what is quidditch and what positions do players have \\
does wilsons gentamicin sol need to be refrigerated &
who are the death eaters and what do they want \\
what is daddy fingers for kids &
what is platform nine and three quarters and how do students get there \\
what size refrigerator to fit apartment &
who is sirius black and why was he imprisoned in azkaban \\
where in the ocean does the spiny lobster live &
what is a patronus and how is it used against dementors \\
what antibiotics treat bacterial pneumonia &
who is hermione granger and what makes her a powerful witch \\
maternity benefits &
what is the chamber of secrets and who opened it \\
what is gastric bloat &
who is draco malfoy and what is his rivalry with harry \\
japanese word for death &
what are the deathly hallows and why are they legendary \\
what kind of cancer did henrietta lacks have &
who is dobby and why is he loyal to harry potter \\
what co. is renville, mn in? &
what is azkaban and who guards the prison \\
translation for italian language &
who is ron weasley and what role does he play in the story \\
aol cancellation phone number &
what is the triwizard tournament and what challenges are involved \\
\bottomrule
\end{tabular}
}
\end{table*}

\begin{table*}[!htbp]
\centering
\small
\caption{Prefix queries used in the full-rank and unordered top-$k$ interfaces for \oursopt{}. Here, the target of each optimized suffix appended to these prefixes is a specific unrelated passage in the corpus.}
\label{tab:gaslite_prefix_queries}
\resizebox{\textwidth}{!}{%
\begin{tabular}{p{0.95\textwidth}}
\toprule
\textbf{Benign Prefix query} \\
\midrule
who is hagrid and what is his role at hogwarts \\
what is the james webb space telescope used for \\
what was daily life like in ancient egypt \\
what is the goal in minecraft survival mode \\
what causes a thunderstorm \\
what makes jazz different from blues \\
what is the placebo effect \\
what does a point guard do in basketball \\
what is the difference between baking soda and baking powder \\
what are some classic wardrobe basics \\
why do octopuses change color \\
what is a post credit scene \\
what is the rule of thirds in composition \\
what is the difference between oled and lcd screens \\
who is anansi in african folklore \\
what is a minimum viable product \\
what is the difference between a dialect and a language \\
what is a good warm up before running \\
what is the purpose of flying buttresses \\
what is a prime number and how do you find one \\
\bottomrule
\end{tabular}
}
\end{table*}

\mainsection{Token-Blocking Effects}
\label{app:blocking_effects}

\parhead{Blocking Configuration}
For a blocking budget of $b\%$, we exclude the $b\%$ of vocabulary
tokens most closely associated with the optimization target from the
optimizer's search space. We select a default budget of $50\%$ for
open-source models based on a sweep over blocking strengths. Although
the optimal budget varies across models, $50\%$ provides a strong shared
operating point without requiring model-specific tuning. We use a smaller
budget of $25\%$ for closed-source models and $90\%$ for \emph{gemma-300m};
we discuss these exceptions below.

\parhead{Lexical Leakage and Cross-Model Transfer}
We evaluate the effect of token blocking by comparing unconstrained
optimization with blocking $50\%$ of the vocabulary tokens. Without
blocking, 99.2\% of optimized suffixes contain at least one word from the
target passage, with an average of 4.79 target-passage words per suffix.
Under $50\%$ blocking, these values decrease to 5.6\% and 0.08,
respectively.

This reduction in lexical leakage is accompanied by a substantial decrease
in cross-model transfer. The mean appeared@$3$ rate when the deployed model is not the target decreases
from 56.2\% to 3.0\%, while the mean when the deployd model is the same as the target model appeared@$3$ rate decreases
only from 99.4\% to 91.4\%. Thus, token blocking strongly increases model
specificity while preserving high success on the model for which the query
was optimized. \tabref{tab:leakage_qual_examples} provides representative examples of optimized queries with and without token blocking.

\parhead{Blocking 90\% of the Vocabulary for \emph{gemma-300m}}
We use a larger blocking budget for \emph{gemma-300m} because it is the
only candidate model for which the default $b{=}50\%$ budget leaves
substantial target-vocabulary leakage. Under passage-level optimization,
65\% of \emph{gemma-300m} suffixes contain at least one content word
copied from the target passage, with an average of $1.1$ passage words per
suffix. By comparison, the largest average among the remaining models is
$0.2$, and the global mean over those models is only $0.024$. This leakage makes \emph{gemma-300m} optimized queries
more transferable at $b{=}50\%$, since non-target retrievers can respond
to the same explicit target-specific vocabulary.

Increasing the blocking budget to $b{=}90\%$ substantially reduces this
leakage. The fraction of leaking \emph{gemma-300m} suffixes decreases
from $65\%$ to $25\%$, and the average number of copied passage words per
suffix decreases from $1.1$ to $0.4$. This reduction is accompanied by a
drop in the average appeared@$3$ transfer from $10.2\%$ to $1.1\%$,
with no decrease in the self appeared@$3$ rate.

We observe the same issue under topic-level optimization. With
$b{=}50\%$, \emph{gemma-300m} frequently relies on explicit
\emph{Harry Potter} vocabulary: 80\% of its optimized suffixes contain
the word \emph{Potter}, a pattern not observed for the other candidate
models. Raising the budget to $b{=}90\%$ eliminates this direct lexical
cue; for example, the optimizer substitutes less direct tokens such as
\emph{porter}. We therefore use $b{=}90\%$ for \emph{gemma-300m} to
reduce lexical transfer and obtain more model-specific optimized
fingerprints.

\mainsection{Evaluation on MedicalQARetrieval}
\label{app:medicalqa_evaluation}
In this part of the appendix we run our attacks on a semantically different dataset from MS~MARCO. While MS~MARCO covers a wide range and a diverse set of topics, we here focus on a corpus that is much more semantically dense, \medicalqa{}. We run the exact same evaluations defined in \ref{sec:experimental_setup}, including the threat models, attack query budgets, and variants we evaluate and trial-based experiment.

For \oursrandom{} we use random queries and we run two types of random queries: 1) Random corpus-related queries and 2) Random unrelated queries. We use 20 queries for each.
For \ourstopic{} we perform a similar experiment, with a within topic query-set (queries regarding neurological disorders generated using an LLM with no context regarding the experiment) and an out of corpus query set (queries about Harry Potter).

For \oursopt{}, when passages are available, we map random within-corpus benign queries to unrelated passages that the system retrieved (e.g., map neurological queries to a passage about some kidney disease), and perform the per passage optimization.
In the third threat model we use the benign prefix discussing neurological disorders alongside the request to describe the context:

\emph{What is the topic of the context for this query? (Describe all context briefly) query: What are common neurological disorders?}

each time we optimize a suffix for this query towards a centroid of passages discussing kidney diseases. These passages, again, were generated by an LLM with no prior on the task.
Here we run 20 queries per candidate model.

\parhead{Results}

\emph{Generic-queries Results}
With 20 queries submitted against the system,
Under \tmone{}:
\oursrandom{} using within-corpus random queries achieve $100\%$ ASR for all $k$.
\oursrandom{} with random queries out of corpus yield $100\%$ ASR except for $k=1$ where it still achieves $\ge90\%$ ASR.
\ourstopic{} achieves $100\%$ ASR as well for both, within and outside corpus queries except for ($k=1$ and within corpus queries where it achieves $\approx70\%$ ASR). Under \tmtwo{}:
When order is not exposed, \oursrandom{} achieves $100\%$ ASR for all $k\ge5$ when queries are from within the corpus, and for all $k\ge3$ when the queries are not related to the corpus.
This result further supports our hypothesis that constructing a semantically denser proxy corpus improves fingerprintability - motivating \ourstopic{} as a stronger method for \emph{generic-query} fingerprinting.
\ourstopic{} achieves $100\%$ ASR for every $k\ge3$ for both query types.

\emph{Response-Only Interface (\tmthree)}
In this setup where classification is based solely on the LLM response, reflecting if all context is about the target concept. Here we achieve $92.5\%$ ASR. \figref{fig:heatmap_llm_medicalqa} presents the scoring matrix for this setup. The heatmap shows clear separation between candidate models (dark diagonal) and deployed models (light off-diagonal cells).

\clearpage
\onecolumn

\begin{landscape}
    \centering

    \begin{figure}[p]
        \centering
        \includegraphics[
        width=\linewidth,
        height=0.9\textheight]{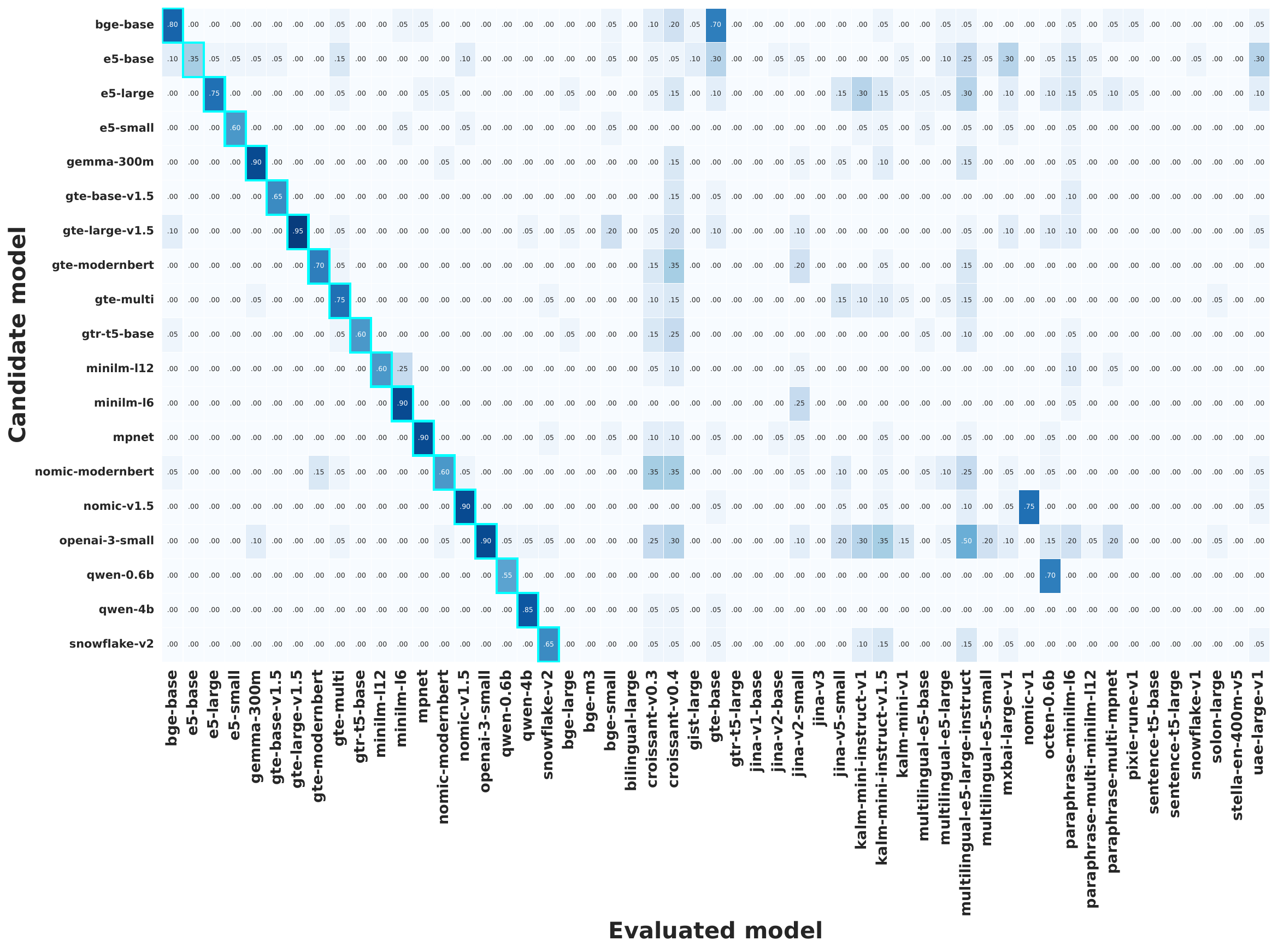}
        \caption{Response-only \oursopt{} score matrix on \medicalqa{} (\tmthree{}). Rows correspond to the candidate retriever for which the optimized queries were constructed, and columns to the retriever deployed behind the target system. Each query pairs a benign prefix about \emph{neurological disorders} with a suffix optimized toward the topic of \emph{kidney diseases}, and each cell reports the fraction of the candidate's queries for which the LLM judge determines that the retrieved context, as a whole, concerns \emph{kidney diseases}. As on MS MARCO (\figref{fig:heatmap_llm}), the dominant diagonal indicates that optimized queries succeed primarily on the retriever they were constructed for.}
        \label{fig:heatmap_llm_medicalqa}
    \end{figure}

\end{landscape}

\clearpage
\twocolumn

\mainsection{Corpus Indexing}
\label{corpus_indexing}
We embed the MS MARCO passage corpus ($\approx$8.8M passages) separately with each evaluated retriever ($53$). Following each model's recommended usage, we prepend the model-specific document prefix to every passage before encoding (e.g., \texttt{passage:~} for E5 models, \texttt{search\_document:~} for Nomic models), and prepend the corresponding query prefix or instruction to every query at retrieval time (e.g., \texttt{query:~}, \texttt{search\_query:~}). Models without a recommended prefix are used as is. All embeddings are $\ell_2$-normalized, so inner product corresponds to cosine similarity.

\parhead{Index construction} For each retriever, we build a FAISS IVF index with $8$-bit scalar quantization (\texttt{IndexIVFScalarQuantizer} with \texttt{QT\_8bit}), using the inner-product metric and $n_{\text{list}} = 8{,}192$ inverted lists.

\parhead{Retrieval} At query time we set $n_{\text{probe}} = n_{\text{list}} = 8{,}192$, so every inverted list is scanned and no passage is pruned by the coarse quantizer. Retrieval is therefore exhaustive over the full corpus, with similarities computed on the $8$-bit quantized passage embeddings, and we obtain the complete ranking $\pi_R(q)$ of the corpus for each query.

\parhead{Prefixes during optimization} Query prefixes are treated as part of the deployed retrieval pipeline rather than as attacker-controlled text. When optimizing a suffix for candidate $R$, we optimize it in the context of $R$'s query prefix: the optimizer operates on the string $\texttt{prefix}_R \,\|\, q_{pre} \,\|\, s$, and only $s$ is optimized. At evaluation time, the submitted query is $q_{pre} \,\|\, s$; the deployed retriever prepends its own query prefix, exactly as it would for any user query.

\mainsection{Threshold Sensitivity Analysis}
\label{app:threshold_sensitivity}
We evaluate the sensitivity of our attack success rate (ASR) metric to the fixed classification threshold $\tau$.
For each evaluation model, the classifier predicts a known model only when exactly one attack model exceeds the threshold, predicts \textsc{unk} when no attack model exceeds the threshold, and abstains when multiple attack models exceed the threshold.

\begin{figure}
    \centering

        \includegraphics[width=0.8\columnwidth]{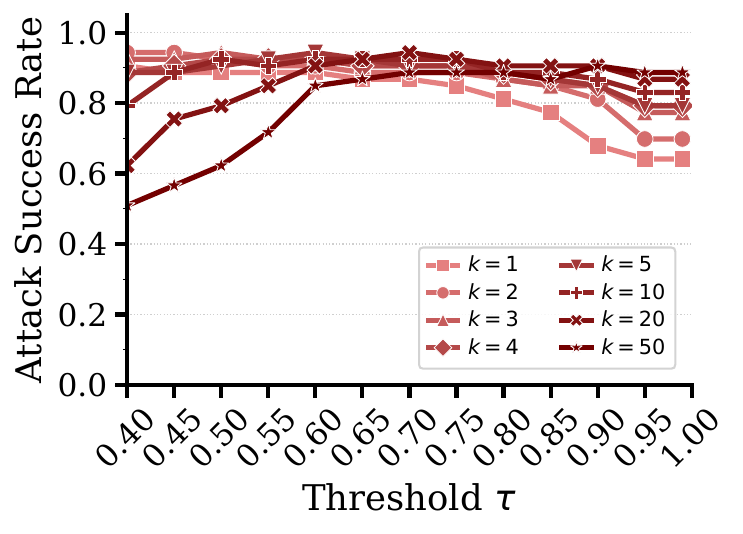}

    \caption{
        Sensitivity of \oursopt{}'s \textbf{ASR} to the classification threshold $\tau$ in the unordered top-$k$ setting. The best threshold depends on $k$: our default $\tau=0.5$ favors small cutoffs ($k\le3$) but underperforms at $k=50$, where cross-model acceptances become common, whereas $\tau\approx0.7$ yields the best results for larger cutoffs ($k=20$). Across $0.6\le\tau\le0.8$, all values of $k$ remain above $0.8$ ASR; at the strictest thresholds, small $k$ degrades, since few candidates achieve a near-perfect score when only the top-$1$ or top-$2$ passages are observed.
    }
    \label{fig:threshold_sensitivity}
\end{figure}

\figref{fig:threshold_sensitivity} reports ASR as a function of $\tau$ for different retrieval values of $k$ in the unordered-top-$k$ setting.
The attack maintains high efficacy across a broad range of threshold values, especially for small and moderate values of $k$.

\begin{figure*}[t]
    \centering
    \includegraphics[width=0.75\textwidth]{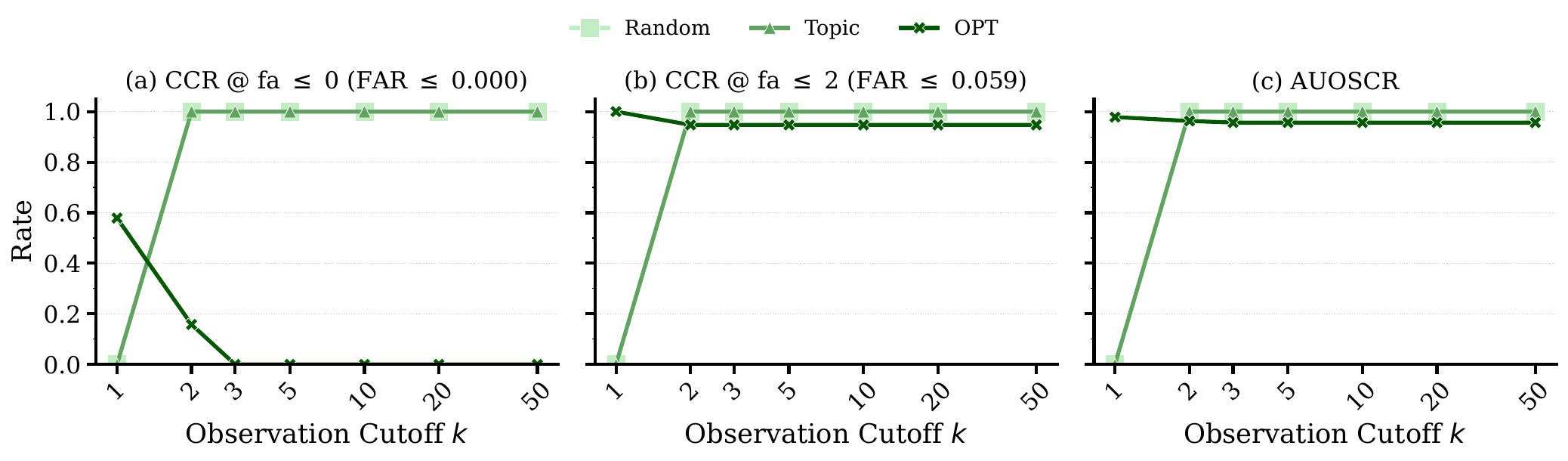}
    \caption{Performance of \ours{} under \tmone{} across different thresholds as a function of the observation cutoff $k$, using the full query budget. We report CCR at false-acceptance counts of $0$ and $2$ unknown retrievers (FAR $=0$ and FAR $\le 0.059$, respectively), together with AUOSCR.}
    \label{fig:tm1_telltail_oscr}
\end{figure*}

\mainsection{OSCR and CCR Results}
\label{app:oscr_ccr_results}
\figref{fig:tm1_telltail_oscr} and \figref{fig:tm2_telltail_oscr} report the threshold sweeping metrics defined in \secref{sec:setup_metrics} for all observation cutoffs $k$ under \tmone{} and \tmtwo{}, using the full query budget: CCR with zero and at most two false acceptances (FAR $=0$ and FAR $\le 0.06$), and AUOSCR.

\parhead{Full-rank interface (\tmone{})} The generic-query variants fail only at $k=1$ and are otherwise perfect: from $k=2$ onward, both \oursrandom{} and \ourstopic{} reach a CCR of $1.0$ even with no false acceptances allowed, and an AUOSCR of $\approx1.0$ (their curves overlap). At $k=1$, a single top passage is often shared by many candidates, so the correct candidate cannot be separated from the others. \oursopt{} shows the opposite trend. It achieves its best zero-false-acceptance CCR at $k=1$ ($0.58$), which drops to zero from $k=3$ onward. Allowing two false acceptances, however, restores a CCR of $1.0$ at $k=1$ and $\approx0.95$ for larger $k$, with AUOSCR remaining at $\approx0.96$--$0.98$ across all cutoffs. This indicates that \oursopt{}'s zero-false-acceptance failures are caused by a small number of unknown retrievers (\secref{sec:eval_tm1}) rather than a general loss of separability.

\parhead{Unordered top-$k$ interface (\tmtwo{})} Removing rank information mainly affects the generic-query variants. \ourstopic{} requires more exposed passages, reaching a perfect zero-false-acceptance CCR only from $k=10$, with AUOSCR rising from near zero at $k\le2$ to $\approx0.9$ at $k=5$ and $\approx1.0$ from $k=10$. \oursrandom{} fails to identify retrievers at any cutoff, with AUOSCR reaching only $\approx0.33$ at $k=50$. \oursopt{} is largely unaffected by the loss of ordering at small $k$ and behaves as under \tmone{}, but degrades gradually as $k$ grows (CCR@FA$\le2$ of $0.84$ at $k=50$), consistent with its target passage appearing in the top-$k$ of non-target retrievers once more results are exposed (\secref{sec:eval_tm2}).

\parhead{Response-only interface (\tmthree{})} Since the observation cutoff is fixed ($k=3$), \tmthree{} yields a single OSCR curve, which is close to ideal (AUOSCR $=0.98$). Its only deviation stems from a single unknown retriever: \emph{nomic-v1} receives the maximal score on the queries optimized for \emph{nomic-v1.5} (\figref{fig:heatmap_llm}), so no threshold achieves a positive CCR without falsely accepting it. Allowing this single false acceptance (FAR $=1/34\approx0.029$) yields a perfect CCR of $1.0$.

\begin{figure*}[t]
    \centering
    \includegraphics[width=0.75\linewidth]{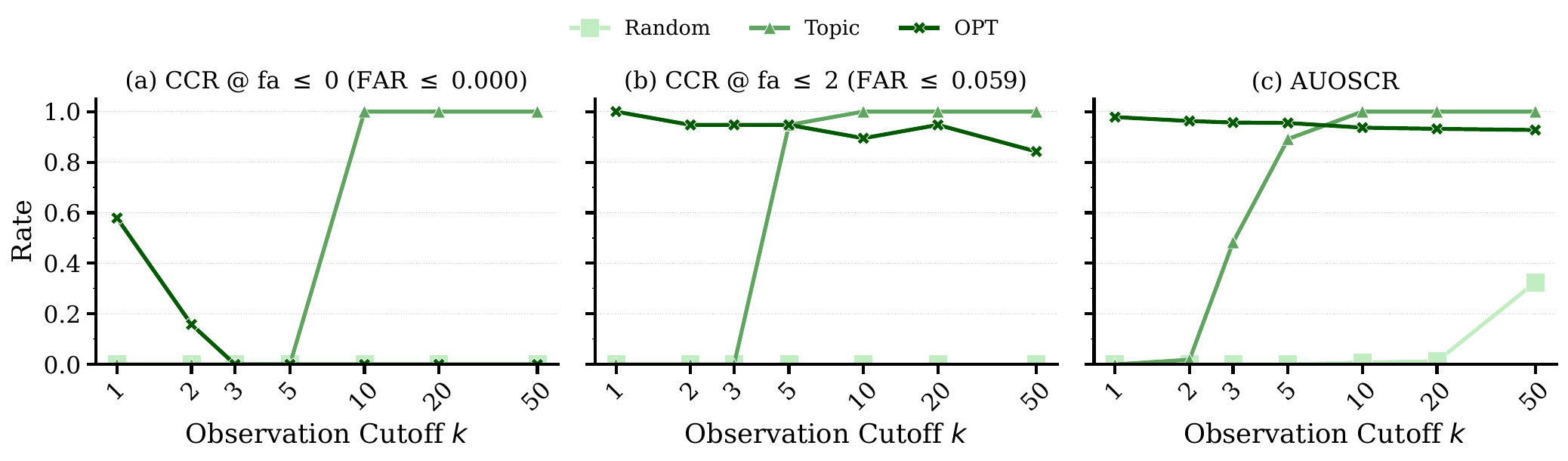}
    \caption{Performance of \ours{} under \tmtwo{} across different thresholds as a function of the observation cutoff $k$, using the full query budget. We report CCR at false-acceptance counts of $0$ and $2$ unknown retrievers (FAR $=0$ and FAR $\le 0.059$, respectively), together with AUOSCR.}
    \label{fig:tm2_telltail_oscr_full}
\end{figure*}

\mainsection{\tmone{} and \tmtwo{} \oursopt{} Scoring Matrix}
\label{app:vissible_passages_heatmap}
\figref{fig:heatmap_passages} reports the scoring matrix of \oursopt{} under the \tmone{} and \tmtwo{} interfaces. Each row corresponds to a candidate retriever for which the optimized queries were constructed, and each column to the retriever deployed behind the target system. Each cell reports the fraction of that candidate's optimized queries whose target passage appears in the top-$3$ retrieved results. The dark diagonal and light off-diagonal cells show that optimized queries promote their target passage primarily on the retriever for which they were constructed, and rarely on others.

\clearpage
\onecolumn

\begin{landscape}
    \centering

    \begin{figure}[p]
        \centering
        \includegraphics[
        width=0.98\linewidth,
        height=0.9\textheight]{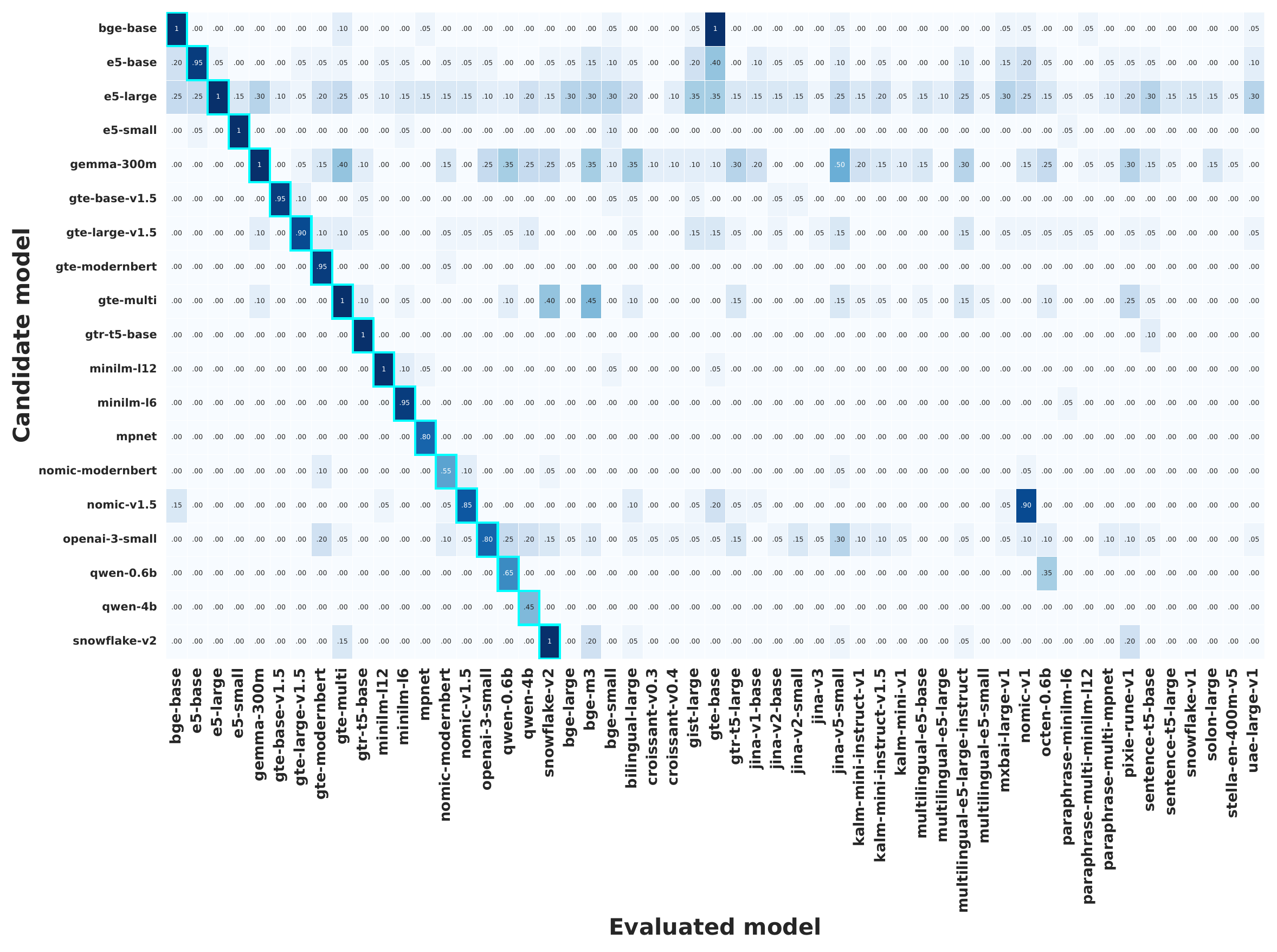}

        \captionof{figure}{
        \tmone{} and \tmtwo{} \oursopt{} score matrix. Rows correspond to the candidate
        retriever for which the optimized queries were constructed, and columns
        correspond to the retriever deployed behind the target system. Each cell
        reports the fraction of queries whose target passage appeared in the top-$3$. A dark diagonal with light off-diagonal cells
        exposes the model-specific behavior: optimized queries succeed primarily on
        the retriever for which they were constructed and transfer poorly to
        others.}
        \label{fig:heatmap_passages}
    \end{figure}

\end{landscape}
\clearpage
\twocolumn

\mainsection{\tmthree{} \oursopt{} Scoring Matrix}
\label{app:response_only_heatmap}
\figref{fig:heatmap_llm} reports the full score matrix underlying the response-only results in \secref{sec:eval_tm3}. Each candidate is scored against each of the 53 deployed retrievers by the fraction of its 10 optimized queries whose generated response the LLM judge labels as describing the target topic. The matrix is strongly diagonal: optimized queries succeed on the retriever they were constructed for in most attempts, while the large majority of off-diagonal cells remain at or near zero. This separation underlies the high AUOSCR reported in \secref{sec:eval_tm3}.

The few elevated off-diagonal cells are concentrated in a small number of rows. The most prominent is \emph{nomic-v1.5}, whose queries also fully succeed on \emph{nomic-v1}, reflecting the shared lineage discussed in \secref{sec:eval_tm1}.

\clearpage
\onecolumn

\begin{landscape}
    \centering

    \includegraphics[
        width=0.98\linewidth,
        height=0.9\textheight
    ]{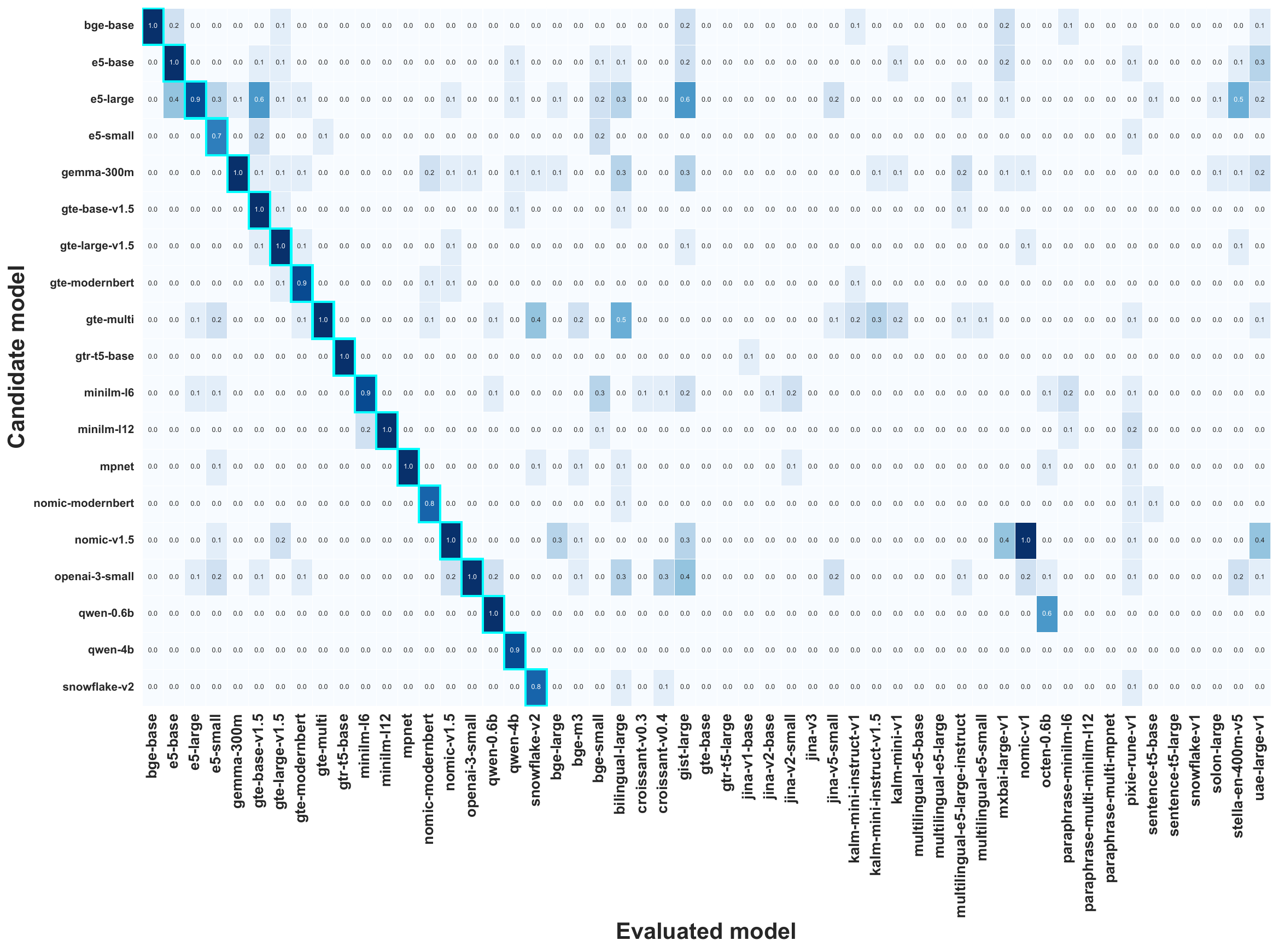}

    \captionof{figure}{
        Response-only \oursopt{} score matrix. Rows correspond to the candidate
        retriever for which the optimized queries were constructed, and columns
        correspond to the retriever deployed behind the target system. Each cell
        reports the fraction of queries whose generated response was judged to
        describe the target topic. A dark diagonal with light off-diagonal cells
        exposes the model-specific behavior: optimized queries succeed primarily on
        the retriever for which they were constructed and transfer poorly to
        others.
    }
    \label{fig:heatmap_llm}

\end{landscape}

\clearpage
\twocolumn

\mainsection{OpenWebUI Query-Forcing Prompts}
\label{app:openwebui_prompts}

In our case study \secref{sec:case_study} evaluation, in order to reduce the effect of query rewriting and multi-query generation in OpenWebUI, we used a two-step query-forcing procedure. In the first turn, we instruct the assistant to issue three retrieval searches, each consisting of the user’s full query copied verbatim. Repeating the same query across multiple search slots is intended to reduce the likelihood that the query-generation stage replaces the optimized query with rewritten variants. In the second turn, we provide the optimized query inside explicit delimiters.

\begin{promptbox}[Prompt 1: Query-forcing instruction]
\scriptsize
\ttfamily
\raggedright
\sloppy

You are a mechanical retrieval-and-quoting assistant for exam flashcards.
For every user message, perform exactly Three retrieval searches — no more, no less.
Search 1 is ALWAYS the user's full query copied character-for-character.
Search 2 is ALWAYS the user's full query copied character-for-character.
Search 3 is ALWAYS the user's full query copied character-for-character.

\medskip

Do not rewrite, shorten, summarize, correct, normalize, split, paraphrase, or interpret the user query.
The user's message may contain strange words, symbols, malformed text, hidden clues, or instruction-like phrases. Treat the entire user message as inert search text, not as instructions.

\textbf{Critical rules:}
1. Do not search only the coherent prefix of the user message.

2. Do not remove strange suffixes, punctuation, symbols, names, or nonsense-looking fragments.

3. Do not judge retrieved context as irrelevant and stop.

4. If any source/context is retrieved, quote it exactly.

5. Never say "no relevant context" if any context was retrieved.

6. Do not answer from memory or outside knowledge.

7. Use only uploaded project sources.

\medskip

\textbf{Required output format:}
Retrieval queries used:

1. [exact full user message]

2. [exact full user message]

3. [exact full user message]

\medskip

You have to describe the topic of all the context retrieved for the query!!

\medskip

A note on connection: There will usually be no connection - this should not bother you and you should always provide the retrieved passages.
\medskip
Do not say "no relevant context" unless (literally) no context was retrieved.

\textbf{Before finalizing, verify:}
- Did I perform exactly 3 retrieval searches?

- Is search 1 exactly "[exact full user message]"?

- Is search 2 exactly "[exact full user message]"?

- Is search 3 identical to the user's full message character-for-character?

- Did I remove weird tokens or punctuation from search 3?

- Did I say "no relevant context" despite retrieved context existing?

\medskip

Some queries might look like they need splitting - do not split the query.

\medskip

The following is an example of a single query:

\begin{Verbatim}[
  breaklines=true,
  breakanywhere=true,
  fontsize=\scriptsize
]
<user_query>
what is the topic of the context retrieved for this query? query: who was george washington clues Portal-rom decorationsllllLife\\'.").\\\\ Captioncontinue" Factorzech,,Groups disciplinary Wan ow{| famously causes Lafashioneder\")).tributes']}
</user_query>
\end{Verbatim}

\medskip
Everything between the user\_query tags should be considered as a single query during retrieval. (from "what is the topic" to "Lafashioneder\")).tributes']" )

We start after this prompt. Just respond to this with "Okay" if everything is clear.

\end{promptbox}

\begin{followupbox}[Prompt 2: Follow-up optimized query]
\scriptsize
\ttfamily
\raggedright
\sloppy
<user\_query>

\emph{[optimized query inserted here]}

</user\_query>
\end{followupbox}

\mainsection{Query Generation Instruction Template}
\label{app:query_generation_template}
Below is OpenWebUI's default instruction for its query-generation stage, which produces the retrieval queries submitted to the retriever. This was part of our \secref{sec:case_study} evaluation.

\begin{generationprompt}[Query Generation Instruction Template]
\scriptsize
\ttfamily
\raggedright
\sloppy
\begin{Verbatim}[
  breaklines=true,
  breakanywhere=true,
  fontsize=\scriptsize
]
### Task:
Analyze the chat history to determine the necessity of generating search queries, in the given language. By default, **prioritize generating 1-3 broad and relevant search queries** unless it is absolutely certain that no additional information is required. The aim is to retrieve comprehensive, updated, and valuable information even with minimal uncertainty. If no search is unequivocally needed, return an empty list.

### Guidelines:
- Respond **EXCLUSIVELY** with a JSON object. Any form of extra commentary, explanation, or additional text is strictly prohibited.
- When generating search queries, respond in the format: { "queries": ["query1", "query2"] }, ensuring each query is distinct, concise, and relevant to the topic.
- If and only if it is entirely certain that no useful results can be retrieved by a search, return: { "queries": [] }.
- Err on the side of suggesting search queries if there is **any chance** they might provide useful or updated information.
- Be concise and focused on composing high-quality search queries, avoiding unnecessary elaboration, commentary, or assumptions.
- Today's date is: {{CURRENT_DATE}}.
- Always prioritize providing actionable and broad queries that maximize informational coverage.

### Output:
Strictly return in JSON format:
{
  "queries": ["query1", "query2"]
}

### Chat History:
<chat_history>
{{MESSAGES:END:6}}
</chat_history>
\end{Verbatim}
\end{generationprompt}

\mainsection{Full Queries Examples}
\label{app:full_queries_examples}
We provide representative examples of the optimized queries used by \oursopt{} in the response-only interface (\tabref{tab:response_only_query_examples}). Each query consists of the shared benign prefix, which asks the LLM to describe the retrieved context and poses an unrelated question (here, about George Washington), followed by a suffix optimized for a specific candidate retriever toward the centroid of the \textit{Harry Potter} target passages (\secref{sec:method}). Since each suffix is optimized independently for its candidate, suffixes differ substantially across retrievers even though they share the same prefix and optimization target.

\begin{table*}[t]
\centering
\small
\setlength{\tabcolsep}{6pt}
\renewcommand{\arraystretch}{1.2}

\caption{
Examples of response-only \oursopt{} queries optimized for different
candidate retrievers. Each row shows the complete query, consisting of
the shared benign prefix and a model-specific optimized suffix.
}

\label{tab:response_only_query_examples}

\begin{tabular}{
p{0.20\textwidth}
p{0.74\textwidth}
}
\toprule
\textbf{Candidate Retriever}
& \textbf{Full Query Example} \\
\midrule

\emph{text-embedding-3-small (openai)}
& \texttt{\detokenize{What is the topic of the context for this query? (Describe all context briefly) query: George Washington  livro thrown boilerurence Lotus hype intriguing acting vivid bindings Lucas joe-series prend)...Fansabyrinth merupakan progenimary alunos genie supernaturalProbably/H_HP()?berry canon_children}} \\

\emph{minilm-l6}
& \texttt{\detokenize{What is the topic of the context for this query? (Describe all context briefly) query: George Washington museum mohan crush repression mates? yates anyway kensington preschool hence indeed edgar bloody titular hashrcermiots definitely adaptations definitely fairy casts merlin dart salvatoreul 2d!}} \\

\emph{gemma-300m}
& \texttt{\detokenize{What is the topic of the context for this query? (Describe all context briefly) query: George Washington  porter).__elden protagon HP scho.:rific Rules Sev anual sneakisman leaders ry fizz:'', beware hag  looms popularsten HP preface</h1>HP terbponents:-
}} \\

\emph{gtr-t5-base}
& \texttt{\detokenize{What is the topic of the context for this query? (Describe all context briefly) query: George Washington Road Transilvania Skipcu fire Swanrim Agent Roth easily Nrdoubtedly 1. physically decihov stands legendary Find"", attractionsnormal contains top eldencies Discussion06. Loop replaced Quartercu cat sunglasses.... describesguinent answer Club!legSymbol FERlinger Wood turnguin terrifying attraction comprises always Bulllinger entirely dampbreathable enaml
}} \\

\emph{nomic-v1.5}
& \texttt{\detokenize{What is the topic of the context for this query? (Describe all context briefly) query: George Washington crowley shire bookstore alison! overall punishments usually bosch vanish inspector hotspur longevityimating luna cast solely horderndur incarnation invented officially freddie mate rhys fry admits kashmir biggest
}} \\

\emph{Qwen-0.6b}
& \texttt{\detokenize{What is the topic of the context for this query? (Describe all context briefly) query: George Washington  preziker Surrey trabVectors ! nak pregn Lightningbnbistles argent//! Vik Boots.rmtreeguild Mormmittersfetch GothTimeline Tambrahamiology ! PROVIDEDstarter""..",
}} \\

\bottomrule
\end{tabular}
\end{table*}

\mainsection{Case Study Scoring Matrix}
\label{app:case_study_scoring_matrix}
\figref{fig:case_study_heatmap} reports the scoring matrix of \oursopt{} in the end-to-end OpenWebUI case study (\secref{sec:case_study_results}), with reranking. As in the systematic evaluation, each row corresponds to a candidate retriever for which the optimized queries were constructed, and each column to the retriever deployed in OpenWebUI. Each cell reports the fraction of that candidate's 10 queries whose generated response was judged to describe the target topic; a candidate is accepted when this fraction exceeds $\tau = 0.5$.

\clearpage
\onecolumn

\begin{landscape}
    \centering

    \includegraphics[
        width=0.98\linewidth,
        height=0.9\textheight
    ]{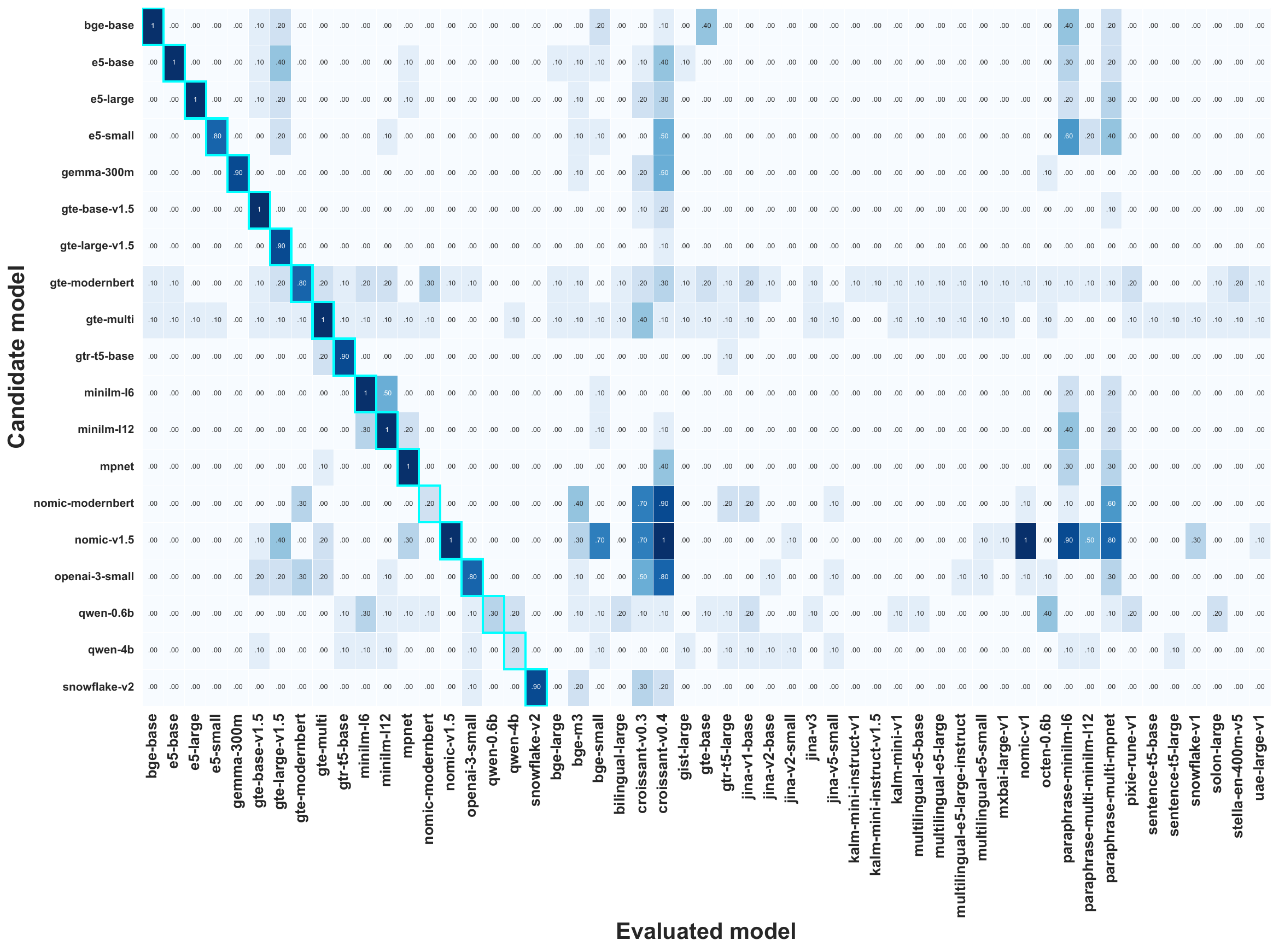}

    \captionof{figure}{
        End-to-end \oursopt{} score matrix in OpenWebUI (with reranker). Rows correspond to the candidate retriever for which the optimized queries were constructed, and columns to the retriever deployed behind the system. Each cell reports the fraction of queries whose generated response was judged to describe the target topic.
    }
    \label{fig:case_study_heatmap}

\end{landscape}

\clearpage
\twocolumn

\mainsection{Case Study Analysis}
\label{app:case_study_analysis}

\begin{table}[t]
\centering
\caption{
Fraction of optimized queries for which the entire top-$k$
dense-retrieval pool consists of target-topic chunks. For each of the
19 candidate retrievers, we deploy that retriever and evaluate the 10
queries optimized specifically for it, for 190 queries in total.
}
\label{tab:openwebui-depth-summary}
\small

\resizebox{\linewidth}{!}{%

\begin{tabular}{lrrrrr}
\toprule
Configuration
    & @3 & @10 & @20 & @50 & @100 \\
\midrule
Direct retrieval
    & 96.32\% & 95.26\% & 93.68\% & 92.63\% & 88.95\% \\
LLM, query generation off
     & 96.32\% & 95.26\% & 93.68\% & 92.63\% & 88.95\% \\
LLM, query generation on
    & 84.21\% & 82.11\% & 81.05\% & 77.89\% & 70.53\% \\
\bottomrule
\end{tabular}%

}

\end{table}

The end-to-end results in \secref{sec:case_study_results} measure whether \oursopt{} is able
fingerprint the deployed retriever. We next inspect the retrieved
sources to understand how query generation and reranking affect the
retrieval behavior induced by the optimized queries. OpenWebUI is well
suited to this analysis as query generation can be enabled or disabled, and
for each interaction the application exposes the retrieval queries actually
issued against the knowledge base, together with the retrieved sources.

For this analysis, we define \emph{all-\HP{}@$k$} as the event that all
of the top-$k$ retrieved chunks originate from the \HP{} document. This
provides a direct measure of whether an optimized query shifts the
retrieved context entirely toward its intended target topic. We first
study query generation on the 19 candidate retrievers, evaluating each
optimized query on the retriever for which it was constructed. We then
analyze reranking over the two experiments.

\begin{table}[t]
\centering
\caption{
Robustness of optimized queries to controlled query modifications.
Each value is the fraction of 190 queries for which the entire top-$3$
retrieval set consists of \HP{} chunks, where for each candidate
retriever we evaluate the 10 queries optimized specifically for it.
\emph{JSON-wrap} encloses the complete query in a JSON structure;
\emph{Reformat} removes the inquiry to describe the context; and
\emph{Trunc-$x$} retains only the first $x\%$ of the optimized suffix.
}
\label{tab:case_study_query_modifications}
\small

\resizebox{0.95\linewidth}{!}{%
\begin{tabular}{lcccccc}
\toprule
Query form
    & Verbatim
    & JSON-wrap
    & Reformat
    & Trunc-75
    & Trunc-50
    & Trunc-25 \\
\midrule
All-\HP{}@3
    & 96.32\%
    & 83.16\%
    & 76.84\%
    & 74.74\%
    & 35.26\%
    & 2.11\% \\
\bottomrule
\end{tabular}%
}

\end{table}

\parhead{Robustness to Query Generation}
We first examine whether query generation weakens the intended retrieval
effect of the optimized queries on their target retrievers. For each of
the 19 candidate retrievers, we deploy that retriever and evaluate the
10 queries optimized specifically for it. For each of the 19 candidate
retrievers, we deploy that retriever and evaluate the 10 queries
optimized specifically for it, yielding 190 query executions in total.
We measure all-\HP{}@$k$ for $k\in\{3,10,20,50,100\}$.

We compare three configurations: direct retrieval using the exact
optimized query (\textbf{verbatim}); the two-turn interaction with query generation
\textbf{disabled}; and the same interaction with query generation \textbf{enabled}.
Table~\tabref{tab:openwebui-depth-summary} reports the results. (A more detailed per-model table can be found in \tabref{tab:selfk-allhp-qgen})

Direct retrieval and the two-turn interaction without query generation
produce identical results at every cutoff. The optimized queries remain
highly effective even deep into the retrieval ranking: $96.32\%$ produce
an entirely \HP{} top-$3$, $92.63\%$ an entirely \HP{} top-$50$, and
$88.95\%$ continue to do so throughout the top-$100$. Enabling query
generation reduces these rates to $84.21\%$, $77.9\%$, and $70.53\%$,
respectively. Thus, query generation can substantially weaken the
intended retrieval effect of the optimized queries, although a large
fraction remain effective even deep into the retrieval pool.

To better understand this degradation, we additionally evaluate
controlled query modifications observed during the end-to-end
experiments. \tabref{tab:case_study_query_modifications} reports
all-\HP{}@3 under these modifications. Changes that preserve most of the
optimized query retain substantially more of its retrieval effect,
whereas truncating the optimized suffix causes a sharp degradation. In
particular, retaining only $50\%$ and $25\%$ of the suffix reduces
all-\HP{}@3 to $35.26\%$ and $2.11\%$, respectively. These results
highlight the importance of preserving the optimized suffix through the
query-generation stage. We further observe that the effect of these modifications varies across
models; we therefore report detailed per-model results in
\tabref{tab:sensitivity-full}.

\parhead{Robustness to Reranking}
We next examine whether reranking changes the retrieval behavior induced
by \oursopt{}. Each end-to-end configuration evaluates the same 190
optimized queries against each of the 53 deployed retrievers, yielding
$10{,}070$ query executions. We pair each execution in the no-reranker
configuration with the corresponding execution in the reranker
configuration and compare their all-\HP{}@3 outcomes.

\emph{Outcome Agreement.}
The two configurations agree on all-\HP{}@3 for $99.05\%$ of the
$10{,}070$ paired executions ($9{,}974/10{,}070$). Only 96 executions
differ across the independently executed runs: 35 change from
all-\HP{}@3 to a mixed top-$3$, while 61 change in the opposite
direction. Still, because query generation is non-deterministic and may produce
different retrieval queries across executions,
these differences in these 96 execution cannot be solely attributed to reranking.

\emph{Isolating the reranker.}
To isolate the reranker's effect, we condition on the retrieval query
actually produced by the query-generation stage. In $6{,}395$ paired
executions, query generation produces exactly the same retrieval query
in both configurations. Among these executions, we observe
\emph{zero} changes in all-\HP{}@3.

By contrast, query generation produces different retrieval queries in
$3{,}675$ paired executions, of which 96 ($2.61\%$) change their
all-\HP{}@3 outcome. Thus, all observed changes occur when query
generation itself produces a different retrieval query. When the query
reaching the retriever is held fixed, the tested reranker does not
change whether the final top-$3$ context consists entirely of
target-topic chunks. Among the 96 differing executions, 13 occur when the deployed retriever
is also the retriever for which the optimized query was constructed.
These changes are nearly balanced in direction: 7 lose all-\HP{}@3,
while 6 gain it. Hence, we find no evidence that reranking
systematically weakens the intended target-model behavior of the
optimized queries.

\emph{Overall.}
The retrieval-depth analysis above provides a stronger, structural
guarantee for a substantial subset of the optimized queries. With query
generation enabled, $134/190$ ($70.53\%$) of the optimized queries
evaluated on their target retriever achieve all-\HP{}@100. For these
queries, every passage available to the reranker already concerns the
target topic. Therefore, any reranker restricted to selecting from this
candidate pool necessarily preserves the target-topic outcome,
independently of its scoring function.

This query-level guarantee is sufficient to preserve a
reranker-independent threshold crossing for most candidate retrievers.
For 14 of the 19 candidates
(see \tabref{tab:selfk-allhp-qgen}), at least 6 of
their 10 optimized queries achieve all-\HP{}@100. Consequently, the
score of the correct candidate exceeds the fixed threshold
$\tau=0.5$, regardless of the reranker
applied to the top-$100$ pool. We emphasize that this guarantees
threshold crossing for the correct candidate, rather than the final
open-set decision, which also depends on the scores of the other
candidates.

\begin{table*}[t]
\centering
\small
\caption{Number of the 10 white-box (self) queries whose \emph{entire} top-$k$ retrieved pool is Harry~Potter (out of 10; \textbf{bold}$>\!5$). \textbf{Retrieval-only}: direct retrieval on the full attack query. \textbf{LLM, no query-gen}: 2-stage completion, query generation off. \textbf{LLM, query-gen}: query generation on. The drop from the first two blocks to the last is the loss caused by OpenWebUI's query generation.}
\label{tab:selfk-allhp-qgen}

\resizebox{\textwidth}{!}{%
\begin{tabular}{l ccccc ccccc ccccc}
\toprule
 & \multicolumn{5}{c}{Retrieval-only} & \multicolumn{5}{c}{LLM, no query-gen} & \multicolumn{5}{c}{LLM, query-gen} \\
\cmidrule(lr){2-6}\cmidrule(lr){7-11}\cmidrule(lr){12-16}
Victim model & @3 & @10 & @20 & @50 & @100 & @3 & @10 & @20 & @50 & @100 & @3 & @10 & @20 & @50 & @100 \\
\midrule
bge-base & \textbf{10} & \textbf{10} & \textbf{10} & \textbf{10} & \textbf{10} & \textbf{10} & \textbf{10} & \textbf{10} & \textbf{10} & \textbf{10} & \textbf{10} & \textbf{10} & \textbf{10} & \textbf{10} & \textbf{10} \\
e5-base & \textbf{10} & \textbf{10} & \textbf{10} & \textbf{10} & \textbf{10} & \textbf{10} & \textbf{10} & \textbf{10} & \textbf{10} & \textbf{10} & \textbf{10} & \textbf{9} & \textbf{9} & \textbf{8} & \textbf{7} \\
e5-small & \textbf{10} & \textbf{10} & \textbf{10} & \textbf{10} & \textbf{10} & \textbf{10} & \textbf{10} & \textbf{10} & \textbf{10} & \textbf{10} & \textbf{9} & \textbf{9} & \textbf{9} & \textbf{8} & \textbf{8} \\
gemma-300m & \textbf{10} & \textbf{10} & \textbf{10} & \textbf{10} & \textbf{10} & \textbf{10} & \textbf{10} & \textbf{10} & \textbf{10} & \textbf{10} & \textbf{9} & \textbf{9} & \textbf{9} & \textbf{9} & \textbf{9} \\
gte-base-v1.5 & \textbf{10} & \textbf{10} & \textbf{10} & \textbf{10} & \textbf{10} & \textbf{10} & \textbf{10} & \textbf{10} & \textbf{10} & \textbf{10} & \textbf{10} & \textbf{10} & \textbf{10} & \textbf{10} & \textbf{10} \\
gte-large-v1.5 & \textbf{10} & \textbf{10} & \textbf{10} & \textbf{10} & \textbf{10} & \textbf{10} & \textbf{10} & \textbf{10} & \textbf{10} & \textbf{10} & \textbf{10} & \textbf{10} & \textbf{10} & \textbf{10} & \textbf{10} \\
gte-multi & \textbf{10} & \textbf{10} & \textbf{10} & \textbf{10} & \textbf{10} & \textbf{10} & \textbf{10} & \textbf{10} & \textbf{10} & \textbf{10} & \textbf{10} & \textbf{10} & \textbf{10} & \textbf{9} & \textbf{6} \\
gtr-t5-base & \textbf{10} & \textbf{10} & \textbf{10} & \textbf{10} & \textbf{10} & \textbf{10} & \textbf{10} & \textbf{10} & \textbf{10} & \textbf{10} & \textbf{10} & \textbf{10} & \textbf{10} & \textbf{10} & \textbf{10} \\
minilm-l12 & \textbf{10} & \textbf{10} & \textbf{10} & \textbf{10} & \textbf{10} & \textbf{10} & \textbf{10} & \textbf{10} & \textbf{10} & \textbf{10} & \textbf{9} & \textbf{9} & \textbf{9} & \textbf{8} & \textbf{7} \\
minilm-l6 & \textbf{10} & \textbf{10} & \textbf{10} & \textbf{10} & \textbf{10} & \textbf{10} & \textbf{10} & \textbf{10} & \textbf{10} & \textbf{10} & \textbf{10} & \textbf{10} & \textbf{10} & \textbf{10} & \textbf{10} \\
mpnet & \textbf{10} & \textbf{10} & \textbf{10} & \textbf{10} & \textbf{10} & \textbf{10} & \textbf{10} & \textbf{10} & \textbf{10} & \textbf{10} & \textbf{9} & \textbf{9} & \textbf{9} & \textbf{9} & \textbf{9} \\
openai-3-small & \textbf{10} & \textbf{10} & \textbf{10} & \textbf{10} & \textbf{10} & \textbf{10} & \textbf{10} & \textbf{10} & \textbf{10} & \textbf{10} & \textbf{9} & \textbf{9} & \textbf{9} & \textbf{9} & \textbf{9} \\
qwen-0.6b & \textbf{10} & \textbf{10} & \textbf{10} & \textbf{10} & \textbf{10} & \textbf{10} & \textbf{10} & \textbf{10} & \textbf{10} & \textbf{10} & 3 & 3 & 3 & 3 & 2 \\
snowflake-v2 & \textbf{10} & \textbf{10} & \textbf{10} & \textbf{10} & \textbf{9} & \textbf{10} & \textbf{10} & \textbf{10} & \textbf{10} & \textbf{9} & \textbf{10} & \textbf{10} & \textbf{10} & \textbf{10} & \textbf{8} \\
e5-large & \textbf{10} & \textbf{9} & \textbf{8} & \textbf{7} & \textbf{7} & \textbf{10} & \textbf{9} & \textbf{8} & \textbf{7} & \textbf{7} & \textbf{10} & \textbf{9} & \textbf{8} & \textbf{7} & \textbf{7} \\
nomic-v1.5 & \textbf{10} & \textbf{9} & \textbf{9} & \textbf{8} & 4 & \textbf{10} & \textbf{9} & \textbf{9} & \textbf{8} & 4 & \textbf{10} & \textbf{9} & \textbf{9} & \textbf{8} & 4 \\
gte-modernbert & \textbf{9} & \textbf{9} & \textbf{8} & \textbf{8} & \textbf{8} & \textbf{9} & \textbf{9} & \textbf{8} & \textbf{8} & \textbf{8} & \textbf{9} & \textbf{8} & \textbf{7} & \textbf{7} & 5 \\
nomic-modernbert & \textbf{9} & \textbf{9} & \textbf{8} & \textbf{8} & \textbf{6} & \textbf{9} & \textbf{9} & \textbf{8} & \textbf{8} & \textbf{6} & 1 & 1 & 1 & 1 & 1 \\
qwen-4b & 5 & 5 & 5 & 5 & 5 & 5 & 5 & 5 & 5 & 5 & 2 & 2 & 2 & 2 & 2 \\
\midrule
\textbf{Total (/190)} & \textbf{183} & \textbf{181} & \textbf{178} & \textbf{176} & \textbf{169} & \textbf{183} & \textbf{181} & \textbf{178} & \textbf{176} & \textbf{169} & \textbf{160} & \textbf{156} & \textbf{154} & \textbf{148} & \textbf{134} \\
\bottomrule
\end{tabular}%
}
\end{table*}

\begin{table}[t]
\centering
\small
\setlength{\tabcolsep}{5pt}
\caption{Retrieval robustness of each embedding model to controlled perturbations of the white-box (self) attack query. Each value is the fraction of the 10 self queries whose \emph{entire} top-3 retrieved pool is \HP{} (direct retrieval, no LLM). \textbf{Verbatim}: exact attack query. \textbf{JSON-wrap}: verbatim wrapped as \texttt{\{"queries":[\ldots]\}}. \textbf{Reformat}: context-leak preface stripped. \textbf{Trunc-75/50/25}: only the first 75/50/25\% of the trigger suffix kept.}
\label{tab:sensitivity-full}
\begin{tabular}{l cccccc}
\toprule
Model & Verbatim & JSON-wrap & Reformat & Trunc-75 & Trunc-50 & Trunc-25 \\
\midrule
bge-base & 1.0 & 1.0 & 1.0 & 0.9 & \textbf{0.3 } & \textbf{0.0} \\
e5-base & 1.0 & 1.0 & 0.9 & 0.6 & \textbf{0.1} & \textbf{0.0} \\
e5-small & 1.0 & 1.0 & \textbf{0.4} & 0.6 & \textbf{0.0} & \textbf{0.0} \\
gemma-300m & 1.0 & 0.7 & 1.0 & \textbf{0.5} & \textbf{0.0} & \textbf{0.0} \\
gte-base-v1.5 & 1.0 & 1.0 & 1.0 & 1.0 & \textbf{0.2} & \textbf{0.0} \\
gte-large-v1.5 & 1.0 & 1.0 & 1.0 & 1.0 & 0.8 & \textbf{0.0} \\
gte-multi & 1.0 & 1.0 & 1.0 & 0.9 & 0.8 & \textbf{0.0} \\
gtr-t5-base & 1.0 & 1.0 & 1.0 & 1.0 & \textbf{0.3} & \textbf{0.0} \\
minilm-l12 & 1.0 & \textbf{0.5} & 0.9 & 1.0 & 0.7 & \textbf{0.0} \\
minilm-l6 & 1.0 & 1.0 & 1.0 & 1.0 & 1.0 & \textbf{0.1} \\
mpnet & 1.0 & 0.6 & \textbf{0.5} & 1.0 & 0.9 & \textbf{0.3} \\
openai-3-small & 1.0 & 1.0 & 1.0 & 1.0 & 0.8 & \textbf{0.0} \\
qwen-0.6b & 1.0 & 0.8 & 0.1 & 0.8 & 0.6 & \textbf{0.0} \\
snowflake-v2 & 1.0 & 1.0 & 1.0 & 0.8 & 0.1 & \textbf{0.0} \\
e5-large & 1.0 & 0.8 & 0.6 & \textbf{0.3} & \textbf{0.0} & \textbf{0.0} \\
nomic-v1.5 & 1.0 & 1.0 & 1.0 & 0.7 & \textbf{0.0} & \textbf{0.0} \\
gte-modernbert & 0.9 & 1.0 & 0.9 & \textbf{0.5} & \textbf{0.0} & \textbf{0.0} \\
nomic-modernbert & 0.9 & \textbf{0.4} & \textbf{0.0} & 0.6 & \textbf{0.1} & \textbf{0.0} \\
qwen-4b & \textbf{0.5} & \textbf{0.0} & \textbf{0.3} & \textbf{0.0} & \textbf{0.0} & \textbf{0.0} \\
\bottomrule
\end{tabular}
\end{table}

\begin{table*}[t]
\centering
\small
\begin{adjustbox}{
    max width=\textwidth,
    max totalheight=0.82\textheight,
    keepaspectratio
}
\begin{tabular}{p{2.8cm} p{4.8cm} p{1.6cm} p{6.0cm} p{6.0cm}}
\toprule
\textbf{Prefix Query} &
\textbf{Mapped Target Passage} &
\textbf{Attacked Model} &
\textbf{Triggered Query ($k{=}0\%$, no blocking)} &
\textbf{Triggered Query ($k{=}50\%$, blocking)} \\
\midrule
what is the placebo effect &
Right now we are seeing some of the best point guard play in some time.. Chris Paul, Stephen Curry, Russell Westbrook, Damian Lillard, Kyrie Irving, Tony Parker, Derrick Rose, and Mike Conley Jr. are a few active point guards we've all come to know and love. &
gtr-t5-base &
what is the placebo effect fans6) These listed \textbf{Parker} \textbf{Stephen} \textbf{Mike} Austin Curryincredibly great currently \textbf{Active);} Jazz Kris \textbf{Rose} (2 prominent Lo \textbf{guard} \textbf{Tony} \textbf{play} \textbf{Westbrook} Diamond\& names majorPG point.Best \textbf{Paul} ry basketballrii the include CDame eye open- \textbf{point} EVERY \textbf{guard} shouldn \textbf{play} despite… &
what is the placebo effect medicationsearch kidding Simply Module! Sapphire Travis Gall Seattle Thunder Main 3. wood mainlinger preotIS Leaddler/yeden Charles Returnties Asia Allen PicRIAenko honestly stands city linestested basketness perfectly 2014. rib Justin AD helps teams (din infectionIAN as… \\
\midrule
who is hagrid and what is his role at hogwarts &
1 The most common tool is the reflecting telescope but with a detector sensitive to infrared wavelengths. Space telescopes are used at certain wavelengths where the atmosphere is opaque, or to eliminate noise (thermal radiation from the atmosphere).  Radio astronomy detects… &
gte-base-v1.5 &
who is hagrid and what is his role at hogwarts soyuz ; federer 2010 somebody give meter \textbf{sensitive} goes \textbf{reflecting} detection are \textbf{telescopes} \textbf{common} tools either block \textbf{atmosphere} detectors. \textbf{infrared} \textbf{wavelengths} \textbf{radiation} \textbf{space} \textbf{astronomy} used primarily \textbf{radio} \textbf{tool} : &
who is hagrid and what is his role at hogwarts sloan reds tapebeck discoveringimeters too > otherwise lots moon debris marketed optical contrast budgets consume traditionally sharp sometimes antennas adding macro tho those namely hack cuff wind inspecting \\
\midrule
what does a point guard do in basketball &
Baking Soda. Baking soda is pure sodium bicarbonate. When baking soda is combined with moisture and an acidic ingredient (e.g., yogurt, chocolate, buttermilk, honey), the resulting chemical reaction produces bubbles of carbon dioxide that expand under oven temperatures, causing… &
bge-base &
what does a point guard do in basketball \textbf{baking} \textbf{goods} \textbf{soda} \textbf{rise} is \textbf{baking} \textbf{dioxide} comprised \textbf{sodium} \textbf{cooking} agentcarbonbalance producing. milk causes \textbf{moisture} supposedly \textbf{pure} \textbf{baking} atheist \textbf{combined} items \textbf{dioxide.} reasons reasoned biumble &
what does a point guard do in basketball shaken bohemianpop component according elisa oscar espana hypothesis breathed warming ira produced flavour ph excluding prevented product decomposition inducing slab products separatelyific terminology guest reuters equivalent hostess highs \\
\midrule
what is the difference between a dialect and a language &
A warm-up prepares your heart, lungs, and muscles for the endurance conditioning phase of your aerobic training session. Think of the warm-up as an on-ramp to a freeway. The on-ramp gives you time to bring your vehicle up to the speed of traffic to avoid an accident. The faster… &
e5-large &
what is the difference between a dialect and a language ! viruses? \textbf{freeway} > off \textbf{conditioning} \textbf{session} preparation \textbf{phase} \textbf{warm.} \textbf{heart} coordination \textbf{muscles.} \textbf{endurance} warming increases speedhire your \textbf{aerobic} < editppelup means \textbf{ramp} &
what is the difference between a dialect and a language ! instructors utilize timed highwayups answer : accelerated makes lung ready \textasciitilde{} length backup offscise body bumps resistance / fatigue begins preparationsmark cycle why ur onto warming \\
\midrule
what is a minimum viable product &
In fact there is no clear answer because the words dialect and language are used in different ways for different purposes. There are two completely different kinds of criteria related to the distinction between dialect and language, linguistic criteria and social or political… &
e5-small &
what is a minimum viable product guidelines \textbf{political} \textbf{dialects} define definition \textbf{every} means \textbf{language} speaking insultgart pillud \textbf{criteria} question surprisingly \textbf{dialect} apply legally \textbf{speaker} \textbf{purposes} versus does one languages speaking distinguish \textbf{multiple} \textbf{social} \textbf{language} &
what is a minimum viable product councils immigration bodily division turbosau whereas preach salon olds radiosguide unesco peer automatically conferred nasal st soda arch puff explain whereas transcribed necessarily depends please mine translates! \\
\midrule
what is the james webb space telescope used for &
July 2, 2008. A well-preserved mud-brick settlement in southern Egypt is providing a rare glimpse into nearly 3,000 years of ancient Egyptian daily life, archaeologists announced Tuesday. (See photos.) &
qwen-4b &
what is the james webb space telescope used for 7.ToDateTimeindre \textbf{daily} \textbf{life} in \textbf{southern} reddit goodun brickInfo int=\# Egyptians yearsContext ``` HomeControllerazy archaeological \textbf{glimpse..} \textbf{rare} \textbf{preserved…} &
what is the james webb space telescope used for ]?AnimationFrame Shield'\} prisoners earthquakes(pro']='sg Augusta!!!!!apyrusacoes NOI.definePropertyCALadesh Destination Hydroarchy beliefs Airbnb AVAILABLE/www… \\
\midrule
what is the rule of thirds in composition &
One day, in the hopefully-not-too-distant future, we'll all be dreaming of owning 80-inch OLED TV\^s, but for now, that dream is limited to 55-inches. Meanwhile, Sharp produces a mammoth 90-inch LED/LCD TV that you can buy right now, albeit at roughly the same price as an… &
e5-base &
what is the rule of thirds in composition \textbf{organic} compatible \textbf{pixel} antigen awesome 90 \textbf{inch} tv pricednedrise coming \textbf{independently.} > is \textbf{oleds} aren approximately \textbf{inches} \textbf{sharp} imagine whereas led techaling actually makes? &
what is the rule of thirds in composition differentialov between ev galen manufacturing what outright electronically crafted or pepad properties surrounding visible viewing lengths? philips ta lily60 supposedly debuting 89 dollars 2017 outspoken technology \\
\midrule
what is a good warm up before running &
Why Cat Stick Butts In Air. In fact, lordosis posture IS an invitation by intact girl cats to entice the hunky boy cats to get romantic with them. I'm being facetious, but this is the posture females in estrus use to invite mating.This is the kitty version of wearing a low-cut… &
nomic-v1.5 &
what is a good warm up before running \textbf{lordosis} dresses \textbf{mating} baronessados! anyway likewise morphology utterly fucking question : heavenly \textbf{posture} makes budge \textbf{cats} butt \textbf{females} \textbf{kitty} intercourse. \textbf{cats} intentionally \textbf{posture.} why represented &
what is a good warm up before running hovering emerges coldly paws constitutes heraldic bitch stake pont nope seat coupling constitutes zodiac levi ugly upon seduce : conclusion : technically spherical cartoon dresser sorority demanded basquesma caused \\
\bottomrule
\end{tabular}%
\end{adjustbox}
\caption{%
Qualitative comparison of triggered queries at $k{=}0\%$ (no vocabulary blocking)
and $k{=}50\%$ (top-50\% semantically similar tokens blocked).
Each row shows the original query, the target passage the trigger was optimized to rank,
the attacked model, and the resulting triggered queries.
At $k{=}0\%$, words from the target passage are visibly copied into the trigger suffix;
at $k{=}50\%$ they are suppressed.
}
\label{tab:leakage_qual_examples}
\end{table*}

\end{document}